\documentclass[sigconf,screen]{acmart}

\setcopyright{none}
\acmDOI{}
\acmISBN{}
\acmYear{}
\copyrightyear{}
\acmPrice{}
\acmConference[]{}{}{}

\usepackage{booktabs}
\usepackage{multirow}
\usepackage{listings}
\usepackage{tabularx}
\usepackage[noend]{algpseudocode}
\usepackage[ruled]{algorithm}
\usepackage{amsmath}
\usepackage{amssymb}
\usepackage{xcolor}
\usepackage{url}
\usepackage{microtype}
\usepackage{paralist}
\usepackage{subfigure}
\usepackage[normalem]{ulem}
\usepackage{verbatim}
\usepackage{soul}
\usepackage{threeparttable}
\usepackage{flushend}

\usepackage{graphicx}
\usepackage{tikz}
\usetikzlibrary{positioning,shapes,shadows,arrows,automata,graphs,patterns}
\usetikzlibrary{decorations.pathmorphing,snakes}
\usepackage{tkz-graph}
\usepackage{wrapfig}

\newcommand{\ignore}[1]{}

\newcommand{\textrel}[1]{\textsc{#1}}
\newcommand{\IntAbs}[1]{\textsf{IntAbs}}
\newcommand{\Name}[1]{\textsf{EC-Diff}}

\newtheorem{definition}{Definition}

\algrenewcommand\algorithmicindent{1.0em}%

\newcommand*\circled[1]{\tikz[baseline=(char.base)]{
            \node[shape=circle,draw,inner sep=0.1pt] (char) {#1};}}

\definecolor{mygreen}{rgb}{0,0.6,0}
\definecolor{mygray}{rgb}{0.5,0.5,0.5}
\definecolor{mymauve}{rgb}{0.58,0,0.82}
\definecolor{dkgreen}{rgb}{0,0.6,0}
\definecolor{darkblue}{rgb}{0.0, 0.0, 0.55}
\definecolor{mypink}{rgb}{0.96,0.76,0.76}

\lstdefinelanguage{program}{%
  basicstyle=\tt,
  classoffset=0,
  keywords={var,const,reg,proc,skip,assume,if,then,else,%
    while,do,call,return,post,await,ewait,yield,init,
    let,in,and,or,true,false,
    for,from,to,
    future, touch,
    fork, rfork, join,
    async, finish,returns,
    spawn, sync, inlet,
    eventloop,
    foreach,
    atomic, method, wait, signal,
    thread, client, begin, end, spawn, repeat, times,
  },
  keywordstyle=\bf,
  classoffset=1,
  morekeywords={g,l},
  keywordstyle=\tt,
  classoffset=0,
  basicstyle=\tt,
  commentstyle=\itshape,
  morecomment=[l]{//},
  morecomment=[s]{/*}{*/},
  morecomment=[n]{(*}{*)},
  mathescape=true,
  escapeinside=`'
}

\lstnewenvironment{program}{%
  \lstset{%
    language={program},
    columns=flexible,
    breaklines=true,
    tabsize=4,
  }
}{}

\makeatletter
\let\OldStatex\Statex
\renewcommand{\Statex}[1][3]{%
  \setlength\@tempdima{\algorithmicindent}%
  \OldStatex\hskip\dimexpr#1\@tempdima\relax}
\makeatother

\begin{document}
%
\title{Datalog-based Scalable Semantic Diffing of Concurrent Programs}


\author{Chungha Sung}
\affiliation{%
  \institution{University of Southern California}
  \city{Los Angeles}
  \state{CA}
  \country{USA}
}

\author{Shuvendu Lahiri}
\affiliation{%
  \institution{Microsoft Research}
  \city{Redmond}
  \state{WA}
  \country{USA}
}

\author{Constantin Enea}
\affiliation{%
  \institution{University Paris Diderot}
  \city{Paris}
  \country{France}
}

\author{Chao Wang}
\affiliation{%
  \institution{University of Southern California}
  \city{Los Angeles}
  \state{CA}
  \postcode{90089}  
  \country{USA}
}

\begin{abstract}

When an evolving program is modified to address issues related to
thread synchronization, there is a need to confirm the change is
correct, i.e., it does not introduce unexpected behavior.  However,
manually comparing two programs to identify the semantic difference is
labor intensive and error prone, whereas techniques based on
model checking are computationally expensive.
%
%
%
To fill the gap, we develop a \emph{fast} and \emph{approximate}
static analysis for computing synchronization differences of two
programs.  The method is fast because, instead of relying on
heavy-weight model checking techniques, it leverages a polynomial-time
Datalog-based program analysis framework to compute
\emph{differentiating} data-flow edges,
i.e., edges allowed by one program but not the other.
Although approximation is used our method is sufficiently accurate due
to careful design of the Datalog inference rules and iterative
increase of the required data-flow edges for representing a difference.
We have implemented our method and evaluated it on a large number of
multithreaded C programs to confirm its ability to produce, often
within seconds, the same differences obtained by human; in contrast,
prior techniques based on model checking take minutes or even hours
and thus can be 10x to 1000x slower.
\end{abstract}



\maketitle

\section{Introduction}

When an evolving concurrent program is modified, often times, the
sequential program logic is not changed; instead, the modification
focuses on thread synchronization, e.g., to optimize performance or
remove bugs such as data-races and atomicity violations.  Since concurrency is hard, it is important to ensure the
modification is correct and does not introduce unexpected behavior.
However, manually comparing two programs to identify the semantic
difference is difficult, and the situation is exacerbated in the
presence of thread interactions: changing a single instruction in a
thread may have a ripple effect on many instructions in other threads.
%
%
Although techniques have been proposed to compute the
synchronization difference, e.g., by leveraging model
checkers~\cite{Bouajjani17}, they are expensive for practice use.
For example, comparing two versions of a program with 578 lines
of C code takes half an hour.

To fill the gap, we develop a \emph{fast} and \emph{approximate}
static analysis to compute such differences with the goal of
reducing analysis time from hours or minutes to seconds.
We assume the two programs are closely related versions of an evolving
software where changes are made to address issues related to
thread synchronization as opposed to the sequential computation logic.
Therefore, same as in prior works~\cite{ShashaS88,Bouajjani17}, we focus on 
synchronization differences.
However, our method is orders-of-magnitude faster because instead of
model checking we leverage a polynomial-time declarative program
analysis framework which uses a set of Datalog rules to model and
reason about thread interactions.

The reason why prior techniques are expensive is because they insist
on being \emph{precise}.  Specifically, they either enumerate interleavings or use a model checker to
ensure a semantic difference, represented as a set of
data-flow edges, is allowed by one of the programs but not by the
other.  However, this in general is equivalent to program
verification, which is an undecidable problem~\cite{Ramalingam00}; even in cases where it
is reduced to a decidable problem, the cost of model checking is too
high.
Our insight is that in practice, it is relatively easy for developers
to inspect a \emph{given} difference to determine if it is feasible;
what is not easy and hence requires tool support is a systematic
exploration of behaviors of the two programs to identify all possible
differences in the first place.
Unfortunately, developing such a tool is a non-trivial task; for
example, the naive approach of comparing individual thread
interleavings would not work due to the often
exponential blowup in the number of interleavings.

Our method avoids the problem by being \emph{approximate} in that it
does not enumerate interleavings.  This also means infeasible behaviors
are sometimes included.  However, our approximation is carefully
designed to take into consideration the program semantics most
relevant to thread interaction. Furthermore, the approximation can be
refined by iteratively increasing the number of data-flow edges used
to characterize a synchronization difference.
We shall show through experiments that our \emph{fast}
and \emph{approximate} analysis method does not lead to overly
inaccurate results.  To the contrary, the synchronization differences
reported by our method closely match the ones identified by human.
Compared to the prior technique based on model checking, which often takes
minutes or even hours, our method can be 10x to 1000x faster.

\begin{figure}[!t]
\vspace{1ex}
\centering
\includegraphics[width=.95\linewidth]{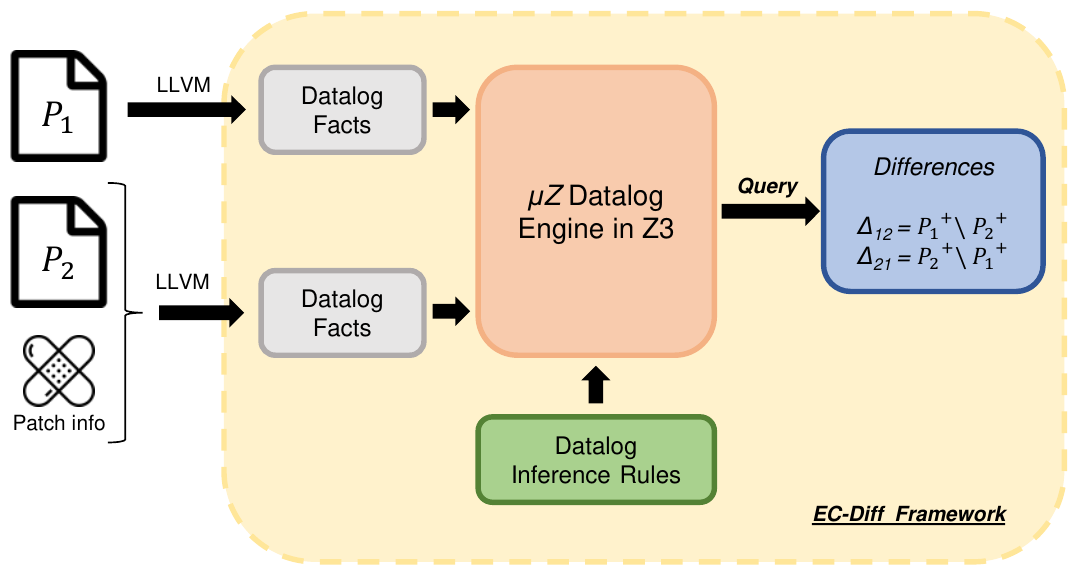}
\vspace{-2ex}
\caption{Overview of our semantic diffing method.}
\label{fig:overview}
\vspace{-2ex}
\end{figure}

Figure~\ref{fig:overview} shows the overall flow of our method.  The
input consists of two versions of a concurrent program: $P_1$ is the
original version, $P_2$ is the changed version, and \emph{patch info}
represents their syntactic difference, e.g., information about which
instructions are added, removed or modified.
The output consists of a set of differences, each of which is
represented by a set of data-flow edges allowed in one of the programs
but not the other.  When data-flow edges are allowed in $P_1$ but not
$P_2$, for example, they represent a removed behavior. 
Conversely, when data-flow edges are allowed in $P_2$ but not $P_1$,
they represent a new behavior introduced by the change.

Our method first generates a set of Datalog facts that encode the
structural information of the control flow graphs.  These facts are
then combined with inference rules that codify the analysis algorithm.
When the combined program is fed to a Datalog solver, the resulting
fixed point contains new relations (facts) that represent the analysis
result.  Specifically, it contains data-flow edges that may occur in
each program.  By comparing data-flow edges from the two programs, we
can identify the semantic differences.

Since program verification is undecidable in general, and with
concurrency, verification is undecidable even for Boolean
programs~\cite{Ramalingam00}, approximation is inevitable.
Our method makes two types of approximations.  The first one is in
checking the feasibility of data-flow edges. 
The second one is related to the number of data-flow edges used to
characterize a difference, also referred to as the \emph{rank} of an
analysis~\cite{Bouajjani17}.  Although in the worst case, a precise
analysis means the rank needs to be as large as the length of the
execution, we restrict it to a small number in our method because
prior research~\cite{MusuvathiQBBNN08,BindalBL13} shows that
concurrency bugs often can be exposed by executions with a bounded
number of context switches.

Since our method is approximate in nature, the usefulness depends on
how close it approaches the ground truth.  Ideally, we want to
have \emph{few} false positives and \emph{few} false negatives.
Toward this end, we choose to stay away from the tradition of
insisting the analysis being either \emph{sound} or \emph{complete}
when one cannot have both.
For a concurrent program, being sound often means \emph{existential}
abstraction: a data-flow edge is considered feasible (in all
interleavings) if it is feasible in an interleaving, and being
complete often means \emph{universal} abstraction: a data-flow edge is
considered feasible only if it is feasible in all interleavings.
Both cases result in extremely coarse-grained approximations, which in
turn lead to numerous false positives or false negatives.
Instead, we want to minimize the difference between our analysis
result and the ground truth.

We have implemented our method in a tool named \Name{}, which uses
LLVM~\cite{Adve03} as the front-end and $\mu Z$~\cite{Hoder11} in Z3
as the Datalog solver.  We evaluated \Name{} on 47 multithreaded
programs with 13,500 lines of C code in total.  These are benchmarks
widely used in prior research~\cite{Beyer15,
Bloem14, WangYGG08, Lu08, Yu09, YangCGW09, Yin11, llvm8441, gcc25530, gcc21334, gcc40518,
gcc3584, glib512624, jetty1187, Herlihy08}: some illustrate real concurrency bug
patterns~\cite{Yu09} and the corresponding patches~\cite{Khoshnood15}
while others are applications from public repositories.
We applied \Name{} to these benchmarks while comparing with the prior
technique of Bouajjani et al.~\cite{Bouajjani17}.  Our results show that \Name{} can detect, often in seconds,
the same differences identified by human.  Furthermore, compared to
the prior technique based on model checking, \Name{} is 10x to 1000x
faster.

To summarize, this paper makes the following contributions:
\begin{itemize}
\item 
We propose a \emph{fast} and \emph{approximate} analysis based on a
polynomial-time declarative program analysis framework to compute
synchronization differences.
\item 
We show why our approximate analysis is reasonably accurate due to the
custom-designed inference rules and iterative increase of the number
of data-flow edges.
\item 
We implement our method in a practical tool and evaluate it on a large
number of benchmarks to confirm its high accuracy and low overhead.
\end{itemize}

The remainder of the paper is as follows.  First, we motivate our work
using examples in Section~\ref{sec:motivation}.  Then, we provide the
technical background in Section~\ref{sec:prelim} before presenting our
analysis method in Section~\ref{sec:constraint}.  This is followed by
our procedures for interpreting the analysis result and optimizing
performance in Section~\ref{sec:optimization}.  We present our
experimental results in Section~\ref{sec:experiment}.  Finally, we
review the related work in Section~\ref{sec:relatedwork} and give our
conclusions in Section~\ref{sec:conclusion}.

\section{Motivation}
\label{sec:motivation}

We use examples to motivate the need for conducting a differential
analysis.
Programs used in these examples illustrate common bug patterns (also
used during our experiments in Section~\ref{sec:experiment}).  In each
example, there are two program versions: the original one may violate
a \emph{hypothetical} assertion and the changed one avoids it.  These
assertions are hypothetical (added for illustration purposes only) in
the sense that our method does not need them to operate.

\subsection{The First Example}

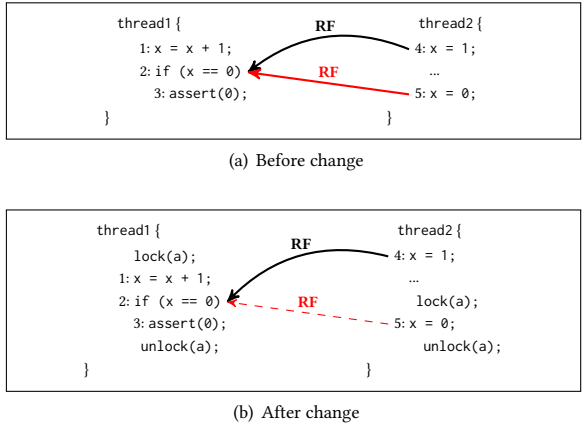
\begin{figure}
\vspace{1ex}
\centering

\subfigure[Before change]{
\centering
\begin{minipage}{.95\linewidth}
\centering
\framebox[.95\linewidth]{
\centering
\begin{minipage}{.95\linewidth}
\centering
\begin{tikzpicture}[node distance=4mm]
  \scriptsize

  \tikzstyle{node}=[minimum size=0pt]
  \tikzstyle{nnode}=[minimum size=0pt,inner sep=0pt]
  \tikzstyle{lnode}=[circle,draw,minimum size=4pt,inner sep=0pt,fill]

  \node[node] (x0)  [] at (0,0) {\texttt{thread1} \{};
  
  \node[node] (x1)  [below=0mm of x0, xshift=5mm] {1:~~\texttt{x = x + 1;}};
    
  \node[node] (x2)  [below=3mm of x0, xshift=5.6mm] {2:~~\texttt{if (x == 0) }};
  
  \node[node] (x3)  [below=6mm of x0, xshift=7mm] {3:~~~~~~~\texttt{assert(0);}};
  
  \node[node] (x4)  [below=9mm of x0, xshift=-5.5mm] {\}};

  \node[node] (y0)  [right=30mm of x0] {\texttt{thread2} \{~~~~~~~};
  
  \node[node] (y1)  [below=0mm of y0, xshift=-0.5mm] {4:~~\texttt{x = 1;}};
    
  \node[node] (y2)  [below=4mm of y0, xshift=-2mm] {...};
  
  \node[node] (y3)  [below=6mm of y0, xshift=-0.5mm] {5:~~\texttt{x = 0;}};
  
  \node[node] (y4)  [below=9mm of y0, xshift=-8mm] {\}};

  \draw[->,>=stealth',thick] (y1.west) to [bend right] node[draw=none,above]{\textbf{RF}} (x2.east);
  \draw[->,>=stealth',thick,color=red] (y3.west) to node[draw=none,above]{\textcolor{red}{\textbf{RF}}} (x2.east);
  
\end{tikzpicture}

\end{minipage}

}
\vspace{1ex}
\end{minipage}

\label{fig:mot1-1}
}

\vspace{1ex}

\subfigure[After change]{
\centering
\begin{minipage}{.95\linewidth}
\centering
\framebox[.95\linewidth]{
\centering
\begin{minipage}{.95\linewidth}
\centering
\begin{tikzpicture}[node distance=4mm]
  \scriptsize

  \tikzstyle{node}=[minimum size=0pt]
  \tikzstyle{nnode}=[minimum size=0pt,inner sep=0pt]
  \tikzstyle{lnode}=[circle,draw,minimum size=4pt,inner sep=0pt,fill]

  \node[node] (x0)  [] at (0,0) {\texttt{thread1} \{};
  
  \node[node] (x5)  [below=0mm of x0, xshift=5mm] {\texttt{lock(a);}};
  
  \node[node] (x1)  [below=3mm of x0, xshift=5mm] {1:~~\texttt{x = x + 1;}};
    
  \node[node] (x2)  [below=6mm of x0, xshift=5.6mm] {2:~~\texttt{if (x == 0) }};
  
  \node[node] (x3)  [below=9mm of x0, xshift=7mm] {3:~~~~~~~\texttt{assert(0);}};
  
  \node[node] (x6)  [below=12mm of x0, xshift=7mm] {\texttt{unlock(a);}};
  
  \node[node] (x4)  [below=15mm of x0, xshift=-5.5mm] {\}};

  \node[node] (y0)  [right=30mm of x0] {\texttt{thread2} \{~~~~~~~};

  \node[node] (y1)  [below=0mm of y0, xshift=-0.5mm] {4:~~\texttt{x = 1;}};
    
  \node[node] (y2)  [below=4mm of y0, xshift=-2mm] {...};
  
  \node[node] (y5)  [below=6mm of y0, xshift=2.5mm] {\texttt{lock(a);}};
  
  \node[node] (y3)  [below=9mm of y0, xshift=-0.5mm] {5:~~\texttt{x = 0;}};
  
  \node[node] (y6)  [below=12mm of y0, xshift=4.5mm] {\texttt{unlock(a);}};
  
  \node[node] (y4)  [below=15mm of y0, xshift=-8mm] {\}};

  \draw[->,>=stealth',thick] (y1.west) to [bend right] node[draw=none,above]{\textbf{RF}} (x2.east);
  \draw[->,>=stealth',color=red, dashed] (y3.west) to node[draw=none,above]{\textcolor{red}{\textbf{RF}}} (x2.east);
  
\end{tikzpicture}

\end{minipage}

}
\vspace{1ex}
\end{minipage}
\label{fig:mot1-2}
}

\vspace{-1ex}
\caption{Example programs with synchronization differences (lock-unlock).}
\label{fig:mot1}
\vspace{-1ex}
\end{figure}

Fig.~\ref{fig:mot1-1} shows a two-threaded program where the shared
variable \texttt{x} is initialized to 0.
The assertion at Line~3 may be violated, e.g., when \texttt{thread1}
executes the statement at Line~2 right after \texttt{thread2} executes
the statement at Line~5.
The reason is because no synchronization operation is used to enforce
any order.

Assume the developer identifies the problem and patches it by adding
locks (Figure~\ref{fig:mot1-2}), the assertion violation will be
avoided.
To see why this is the case, consider the data-flow edge from Line~5
to Line~2: due to the critical sections enforced by lock-unlock pairs,
the load of \texttt{x} at Line~2 is not affected by the store
of \texttt{x} at Line~5.
For example, if the critical section containing Line~5 is executed
first, the subsequent \texttt{unlock(a)} must be executed before
the \texttt{lock(a)} in \texttt{thread1}, which in turn must be
executed before Line~1 and Line~2.  Since the store of \texttt{x} at
Line~1 is the most recent, the load of \texttt{x} at Line~2 will get
its value, not the value written at Line~5.

Thus, the allowed data-flow edges are as follows:
\texttt{RF(L4,L2)} and \texttt{RF(L5,L2)} for the original program, 
and \texttt{RF(L4,L2)} for the changed program.
This notion of comparing concurrent executions was introduced by
Shasha and Snir~\cite{ShashaS88} and extended by Bouajjani et
al.~\cite{Bouajjani17}, although in both cases, enumeration or
model checking techniques were used. 
In our work, the goal is to avoid such heavyweight analyses while
maintaining sufficient accuracy.

In addition to \texttt{RF} edges, there are other types of relations
considered during our analysis, including program order, inter-thread
order imposed by thread \emph{create}, \emph{join}, \emph{signal-wait}
as well as \emph{store-store} order.  
Nevertheless, when interpreting the final results, we focus on
differences in the \texttt{RF} edges because they affect the
externally observable behavior of a program, e.g., characterized by
assertions and other reachability properties.

\subsection{The Second Example}

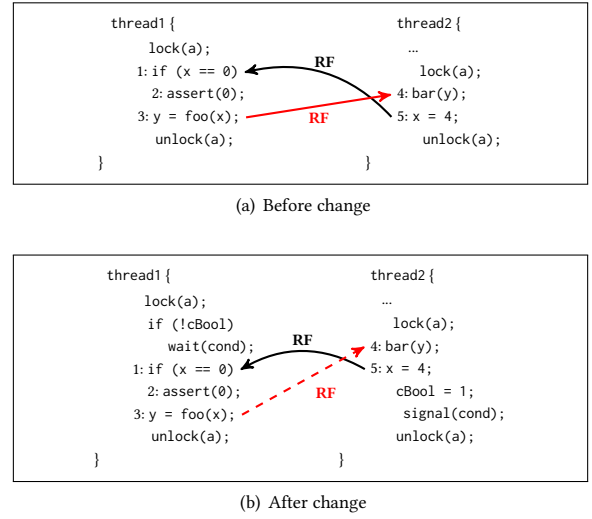
\begin{figure}
\vspace{1ex}
\centering

\subfigure[Before change]{
\centering
\begin{minipage}{.95\linewidth}
\centering
\framebox[.95\linewidth]{
\centering
\begin{minipage}{.95\linewidth}
\centering
\begin{tikzpicture}[node distance=4mm]
  \scriptsize

  \tikzstyle{node}=[minimum size=0pt]
  \tikzstyle{nnode}=[minimum size=0pt,inner sep=0pt]
  \tikzstyle{lnode}=[circle,draw,minimum size=4pt,inner sep=0pt,fill]

  \node[node] (x0)  [] at (0,0) {\texttt{thread1} \{};
  
  \node[node] (x1)  [below=0mm of x0, xshift=5mm] {\texttt{lock(a);}};
  
  \node[node] (x2)  [below=3mm of x0, xshift=6mm] {1:~~\texttt{if (x == 0) }};
  
  \node[node] (x3)  [below=6mm of x0, xshift=7.5mm] {2:~~~~~~~\texttt{assert(0);}};
  
  \node[node] (x4)  [below=9mm of x0, xshift=6.2mm] {3:~~\texttt{y = foo(x);}};
    
  \node[node] (x5)  [below=12mm of x0, xshift=7mm] {\texttt{unlock(a);}};
  
  \node[node] (x6)  [below=15mm of x0, xshift=-5.5mm] {\}};

  \node[node] (y0)  [right=28mm of x0] {\texttt{thread2} \{~~~~~~~};
  
  \node[node] (y1)  [below=1mm of y0, xshift=-2mm] {...};
  
  \node[node] (y2)  [below=3mm of y0, xshift=3mm] {\texttt{lock(a);}};
  
  \node[node] (y3)  [below=6mm of y0, xshift=0.5mm] {4:~~\texttt{bar(y);}};
  
  \node[node] (y4)  [below=9mm of y0, xshift=0mm] {5:~~\texttt{x = 4;}};
  
  \node[node] (y5)  [below=12mm of y0, xshift=4.5mm] {\texttt{unlock(a);}};
  
  \node[node] (y6)  [below=15mm of y0, xshift=-8mm] {\}};

  \draw[->,>=stealth',thick] (y4.west) to [bend right] node[draw=none,above]{\textbf{RF}} (x2.east);
  \draw[->,>=stealth',thick,color=red] (x4.east) to node[draw=none,below]{\textcolor{red}{\textbf{RF}}} (y3.west);
  
\end{tikzpicture}

\end{minipage}

}
\vspace{1ex}
\end{minipage}

\label{fig:mot2-1}
}

\vspace{1ex}

\subfigure[After change]{
\centering
\begin{minipage}{.95\linewidth}
\centering
\framebox[.95\linewidth]{
\centering
\begin{minipage}{.95\linewidth}
\centering
\begin{tikzpicture}[node distance=4mm]
  \scriptsize

  \tikzstyle{node}=[minimum size=0pt]
  \tikzstyle{nnode}=[minimum size=0pt,inner sep=0pt]
  \tikzstyle{lnode}=[circle,draw,minimum size=4pt,inner sep=0pt,fill]

  \node[node] (x0)  [] at (0,0) {\texttt{thread1} \{};
  
  \node[node] (x1)  [below=0mm of x0, xshift=5mm] {\texttt{lock(a);}};
  
  \node[node] (x7)  [below=3mm of x0, xshift=7mm] {\texttt{if (!cBool)}};
  
  \node[node] (x8)  [below=6mm of x0, xshift=9.5mm] {~~~~~~~\texttt{wait(cond);}};
  
  \node[node] (x2)  [below=9mm of x0, xshift=6mm] {1:~~\texttt{if (x == 0) }};
  
  \node[node] (x3)  [below=12mm of x0, xshift=7.5mm] {2:~~~~~~~\texttt{assert(0);}};
  
  \node[node] (x4)  [below=15mm of x0, xshift=6.2mm] {3:~~\texttt{y = foo(x);}};
    
  \node[node] (x5)  [below=18mm of x0, xshift=7mm] {\texttt{unlock(a);}};
  
  \node[node] (x6)  [below=21mm of x0, xshift=-5.5mm] {\}};

  \node[node] (y0)  [right=25mm of x0] {\texttt{thread2} \{~~~~~~~};
  
  \node[node] (y1)  [below=1mm of y0, xshift=-2mm] {...};
  
  \node[node] (y2)  [below=3mm of y0, xshift=3mm] {\texttt{lock(a);}};
  
  \node[node] (y3)  [below=6mm of y0, xshift=0.5mm] {4:~~\texttt{bar(y);}};
  
  \node[node] (y4)  [below=9mm of y0, xshift=0mm] {5:~~\texttt{x = 4;}};
  
  \node[node] (y7)  [below=12mm of y0, xshift=4.5mm] {\texttt{cBool = 1;}};
  
  \node[node] (y8)  [below=15mm of y0, xshift=7mm] {\texttt{signal(cond);}};
  
  \node[node] (y5)  [below=18mm of y0, xshift=4.5mm] {\texttt{unlock(a);}};
  
  \node[node] (y6)  [below=21mm of y0, xshift=-8mm] {\}};

  \draw[->,>=stealth',thick] (y4.west) to [bend right] node[draw=none,above]{\textbf{RF}} (x2.east);
  \draw[->,>=stealth',thick,color=red, dashed] (x4.east) to node[draw=none,below, xshift=3mm]{\textcolor{red}{\textbf{RF}}} (y3.west);
  
\end{tikzpicture}

\end{minipage}

}
\vspace{1ex}
\end{minipage}
\label{fig:mot2-2}
}

\vspace{-1ex}
\caption{Example programs with synchronization differences (signal-wait).}
\label{fig:mot2}
\vspace{-1ex}
\end{figure}

Fig.~\ref{fig:mot2} shows a more sophisticated example: the use of
\emph{signal-wait}, which is often  difficult for static analyzers.   
%
Since the variable \texttt{x} is initialized to 0, when the critical
section in \texttt{thread1} is executed before \texttt{thread2}, the
load of \texttt{x} at Line~1 will get the value 0, which leads to the
assertion violation in Fig.~\ref{fig:mot2-1}.
Assume the intended behavior is for \texttt{thread2} to complete
first, an inter-thread execution order must be enforced, e.g., by
using the \emph{signal-wait} pair shown in Fig.~\ref{fig:mot2-2}.  The
assertion violation is avoided because the load of \texttt{x} at
Line~1 can only read from the store of \texttt{x} at Line~5.

To correctly deploy the \emph{signal-wait} pair, a variable
named \texttt{cBool} needs to be added.  If the operating system
voluntarily schedules \texttt{thread2} first, \texttt{thread1} needs
to be aware -- by checking the value of \texttt{cBool} -- and then
skips the execution of \emph{wait}; otherwise, \emph{wait} may get
stuck because the corresponding \emph{signal} has already been fired
(and lost).  But if \texttt{thread1} is executed first,
since \texttt{cBool} has not been set, it will invoke \emph{wait}
which forces the corresponding \emph{signal} to be sent.

As for the data-flow edges, we can see that \texttt{RF(L5,L1)}
and \texttt{RF(L3,L4)} are allowed in the original program, but
only \texttt{RF(L5,L1)} is allowed in the changed program.  
\texttt{RF(L3,L4)} is not allowed because Line~4 must happen
before Line~5, Line~5 must happen before \emph{signal},
and \emph{signal} must happen before \emph{wait}, which resides before
Lines~1-3 in \emph{thread1}.  Thus, there is a cycle (contradiction).

\subsection{How Our Method Works}

Our method differs from prior techniques which rely on either
enumerating interleavings and conducing pairwise
comparison~\cite{ShashaS88}, or model checking based
techniques~\cite{Bouajjani17}. Both are computationally
expensive.  Instead, we use lightweight static analysis.

Our method represents the control and data dependencies of each
program as a set of Datalog \emph{facts}. We also design a set of
Datalog \emph{inference rules}, which capture our algorithm for
deriving new facts from existing facts.  Leveraging a Datalog solver,
we can repeatedly apply the inference rules over the facts until a
fixed point is reached.  We will explain details of our Datalog facts
and inference rules in Section~\ref{sec:constraint}.

\begin{figure}
\centering

\subfigure{

\begin{minipage}{\linewidth}
\centering
\scalebox{0.8}{

\begin{tabular}{ | c | c | c |}
\hline
 
    & $mustHB$ & $ \{ (1, 2), (2,3), (1, 3), (4, 5) \}$ \\
    \cline{2-3}
    Fig~\ref{fig:mot1-1} & $mayHB$ & $mustHB$ $ \cup $ $ \{(1,4),(1,5),(2,4),$ \\
    &  & $(2,5),(3,4),(3,5),(4,1),(4,2), ... \}$ \\
    \cline{2-3}
     & \textbf{MayRF} & $ \{(4,1), (4,2), (5,1), \textbf{(5,2)} \}$ \\
    \hline
    
    & $mustHB$ & $ \{ (1, 2), (2,3), (1, 3), (4, 5) \}$ \\
    \cline{2-3}
     Fig~\ref{fig:mot1-2} & $mayHB$ & $mustHB$ $ \cup $ $ \{(1,4),(1,5),(2,4),$ \\
    &  & $(2,5),(3,4),(3,5),(4,1),(4,2), ... \}$ \\
    \cline{2-3}
     & \textbf{MayRF}  & $ \{(4,1), (4,2), (5, 1) \}$ \\ 
    \hline

\end{tabular}
}
\\

\vspace{1ex}
\end{minipage}

\label{fig:motTable-1}
}

\subfigure{
\centering

\begin{minipage}{\linewidth}
\centering
\scalebox{0.8}{

 \begin{tabular}{ | c | c | c |}
 \hline
 
    & $mustHB$ & $ \{ (1, 2), (2,3), (1, 3), (4, 5) \}$ \\
    \cline{2-3}
    Fig~\ref{fig:mot2-1} & $mayHB$ & $mustHB$ $ \cup $ $ \{(1,4),(1,5),(2,4),$ \\
    &  & $(2,5),(3,4),(3,5),(4,1),(4,2), ... \}$ \\
    \cline{2-3}
     & \textbf{MayRF} & $ \{\textbf{(3,4)}, (5,1), (5,3) \}$ \\
    \hline
    
    & $mustHB$ & $ \{ (1, 2), (2,3), (1, 3), (4, 5), $  \\
    Fig~\ref{fig:mot2-2} & & $ \textbf{(4,1),(4,2),(4,3),(5,1),(5,2),(5,3)} \}$ \\
    \cline{2-3}
      & $mayHB$ & $mustHB$ \\
    \cline{2-3}
    & \textbf{MayRF}  & $ \{(5,1), (5, 3) \}$ \\
    \hline

\end{tabular}
}

\vspace{1ex}
\end{minipage}
\label{fig:motTable-2}
}

\vspace{-1ex}
\caption{Analysis steps for programs in Figs.~\ref{fig:mot2-1} and \ref{fig:mot2-2}.}
\label{fig:motTable}
\vspace{-1ex}
\end{figure}

For now, consider the steps of computing synchronization differences
for the programs in Fig.~\ref{fig:mot1} and Fig.~\ref{fig:mot2}, which
are outlined by the tables in Fig.~\ref{fig:motTable}.

First, our method computes must-happen-before (\texttt{mustHB}) edges,
which represent the execution order of two instructions respected by
all thread interleavings. 
From \texttt{mustHB}, our method computes may-happen-before
(\texttt{mayHB}) edges, which represent the execution order respected
by some interleavings, e.g., thread context switches not contradicting
to \texttt{mustHB}.
From \texttt{mayHB}, our method computes \texttt{MayRF} edges,
which represent data flows (over shared variables) from store
instructions to the corresponding load instructions.

The \texttt{MayRF} edges are over-approximated in that, if an edge
is included in \texttt{MayRF}, the corresponding data
flow \emph{may} occur in an execution.  But if an edge is not included
in \texttt{MayRF}, we know for sure the corresponding data flow is
definitely infeasible.
For example, in Fig.~\ref{fig:motTable}, \texttt{MayRF} has four
edges for Fig.~\ref{fig:mot1-1} but only three edges for
Fig.~\ref{fig:mot1-2}.  \texttt{RF(L5,L2)} is no longer allowed in the
changed program, indicating it is a difference between the two
programs.

For the example in Fig.~\ref{fig:mot2}, we compute \texttt{mustHB}
based on the sequential program order and, in Fig.~\ref{fig:mot2-2},
the inter-thread execution order imposed by \emph{signal-wait}.
Then, from \texttt{mustHB} we compute \texttt{mayHB}, which includes
edges in \texttt{mustHB} and more.  For Fig.~\ref{fig:mot2-1}, since
there is no restriction on the inter-thread execution order, all pairs
of events are included, whereas for Fig.~\ref{fig:mot2-2}, there is only one-way data flow. 
Finally, we compute \texttt{MayRF} based on \texttt{mayHB}.  There
are three edges for Fig.~\ref{fig:mot2-1} but only two for
Fig.~\ref{fig:mot2-2}.

\subsection{The Rank of an Analysis}

When comparing \texttt{MayRF} in these two examples, we identify the
difference as edges allowed in only one of the two programs, such
as \texttt{RF(L5,L2)}in Fig.~\ref{fig:mot1} and \texttt{RF(L3,L4)} in
Fig.~\ref{fig:mot2}.
%
%

However, even if \texttt{MayRF} edges are allowed individually, they
may not occur in the same execution.
For example, \texttt{RF(L5,L1)} and \texttt{RF(L3,L4)} in
Fig.~\ref{fig:mot2-1} cannot occur together  because,
otherwise, they form a cycle together with the program order
edges.  Our method has inferences rules designed to check if two or
more data-flow edges can occur together---this is referred to as
the \emph{rank}~\cite{Bouajjani17}.

With the notion of rank, we can 
capture ordered sets of \texttt{MayRF} edges, as opposed to
individual \texttt{MayRF} edges. 
Thus, even if the \texttt{MayRF} relation remains the same, there may
be differences of high ranks: two or more edges from \texttt{MayRF}
may occur together in $P_1$ but not in $P_2$.
We will present our method for checking such differences in
Section~\ref{sec:optimization} following the baseline procedure in
Section~\ref{sec:constraint}.

\section{Preliminaries}
\label{sec:prelim}


\subsection{Partial trace comparison}

To compare the synchronizations of two concurrent programs, we use the
notion of partial trace introduced by Shasha and
Snir~\cite{ShashaS88} and extended by Bouajjani et
al.~\cite{Bouajjani17}.
Let $P$ be a program and $\mathbb{G}$ be the set of global variables
shared by threads in $P$.  For each $x\in\mathbb{G}$, let $W(x)$
denote a store instruction and $R(x)$ denotes a load instruction. 
Let $\mathbb{I}$ be the set of all instructions in the program.  Any
binary relation over these instructions is a subset of
$\mathbb{I} \times \mathbb{I}$.

For example, $\hat{so} \subseteq \mathbb{I}\times\mathbb{I}$ is a
relation that orders the store instructions; $W_1(x) < W_2(x)$ means
$W_1 \in \mathbb{I}$~ is executed before $W_2 \in \mathbb{I}$.
Thus, in Fig.~\ref{fig:mot1-1},  \texttt{(L1,L4)}, \texttt{(L4,L1)},
\texttt{(L1,L5)}, \texttt{(L5,L1)}, \texttt{(L4,L5)}
belong to $\hat{so}$, but \texttt{(L5,L4)} does not belong to
$\hat{so}$ because it is not consistent with the program order.

Similarly, $\hat{rf}$ is a relation between load and store
instructions.  In Fig.~\ref{fig:mot1-1}, we have \texttt{(L4,L2)}
and \texttt{(L5,L2)} in $\hat{rf}$, meaning the load at Line~2 may
read from values written at Lines 4 and 5.
Given $\hat{so}$ and $\hat{rf}$, we define $\hat{sets}$ as a set
of \emph{subsets} of $\hat{rf}\cup\hat{so}$, where each element
$ss\in\hat{sets}$ has at most $k$ edges.

Edges in $ss$ are from
either $\hat{rf}$ or $\hat{so}$ -- they capture the abstract trace.
The number $k$, which is called the \emph{rank}~\cite{Bouajjani17}, is
bounded by the length of the trace.

\vspace{1ex}
\begin{definition}[Abstract Trace with Rank $k$]
An abstract trace with rank $k$ is a tuple $\hat{T}
= \langle \hat{so}, \hat{rf}, \hat{sets}, k \rangle$, where
$\hat{so} \subseteq \{ W_1(x) \times W_2(x)~|~ W_1\in\mathbb{I}, W_2\in\mathbb{I}, \mbox{ and } W_1<W_2 \mbox{ in some execution trace} \}$,
$\hat{rf} \subseteq \{ W(x) \times R(x) ~|~ W\in\mathbb{I} \mbox{ and } R\in\mathbb{I} \}$, and
$\hat{sets} \subseteq \{ ss \subseteq \hat{rf} \cup \hat{so} ~|~
|ss| \leq k\}$.
\end{definition}

\vspace{1ex}
Given the abstract traces $\hat{T_1}$ and $\hat{T_2}$ of two programs
$P_1$ and $P_2$, respectively, we define their difference as
$\Delta = (\Delta_{12}, \Delta_{21})$, where $\Delta_{12}
= \hat{T_1} \setminus \hat{T_2}$ and $\Delta_{21}
= \hat{T_2} \setminus \hat{T_1}$.  
Next, we define what it means for $\hat{T_1}$ to be a refinement of
$\hat{T_2}$, denoted $\hat{T_1}\subseteq \hat{T_2}$.

\vspace{1ex}
\begin{definition}[Abstract Trace Refinement]
Given two abstract traces 
$\hat{T_1} = \langle \hat{so_1}, \hat{rf_1}, \hat{sets_1}, k \rangle$ and 
$\hat{T_2} = \langle \hat{so_2}, \hat{rf_2}, \hat{sets_2}, k \rangle$, 
we say $\hat{T_1}$ is a refinement of $\hat{T_2}$, denoted
$\hat{T_1} \subseteq \hat{T_2}$, if and only if 
$\hat{so_1}   \subseteq \hat{so_2}$,
$\hat{rf_1}   \subseteq \hat{rf_2}$, and 
$\hat{sets_1} \subseteq \hat{sets_2}$.
\end{definition}

\vspace{1ex}
That is, when $\hat{T_1} \subseteq \hat{T_2}$, the abstract behavior
of $P_1$ is covered by that of $P_2$.  And the difference
($\hat{T_2}\setminus\hat{T_1}$) is characterized by
$\hat{so_2}\setminus\hat{so_1}$, $\hat{rf_2}\setminus\hat{rf_1}$, and
$\hat{sets_2}\setminus\hat{sets_1}$.
Finally, if the abstract traces
of $P_1$ and $P_2$ refine each other, we say they are \emph{rank-$k$
equivalent}.

Although comparison of abstract traces involves $\hat{so}$ and
$\hat{rf}$, when reporting the differences, we focus on the $\hat{rf}$
edges only because they directly affect the \emph{observable}
behaviors of the programs. In contrast, store-store ordering
($\hat{so}$) may not be observable unless they also affect the
read-from ($\hat{rf}$) edges.

\subsection{Datalog-based Analysis}

Datalog is a logic programming language but in recent years has been
widely used for declarative program analysis~\cite{Dawson96, Whaley04,
Livshits05, Hajiyev06, Heintze01, Bravenboer09, ZhangMGNY14, AlbarghouthiKNS17}.
The main advantage is that a Datalog program is polynomial-time
solvable and the corresponding fixed-point computation maps naturally
to fixed-point computations in program analysis algorithms.
In this context, structural information of the program is represented
as relations called the \emph{facts}, while the fixed-point algorithm
is expressed as recursive relations called the \emph{inference rules}.

Consider a relation named \textrel{PO}$(a, b)$, which represents the
program order of two immediate adjacent instructions $a$ and $b$,
while \textrel{HB}$(c, d)$ means $c$ must happen before $d$.
First, we write down the Datalog facts based on the CFG structure:

\vspace{1ex}
{\footnotesize
$\textrel{PO}(s_1, s_2)$,
$\textrel{PO}(s_1, s_3)$,
$\textrel{PO}(s_2, s_4)$,
$\textrel{PO}(s_3, s_4)$, 
$\textrel{PO}(s_4, s_5)$.
}
\vspace{1ex}

\noindent
Then, we write down the Datalog inference rules:

\vspace{1ex}
{\footnotesize
\raggedright
~~~$\textrel{HB}(a, b) \leftarrow \textrel{PO}(a, b)$

~~~$\textrel{HB}(c, e) \leftarrow \textrel{HB}(c, d) \land \textrel{HB}(d, e) $ 
}
\vspace{1ex}

\noindent
Here, the left arrow ($\leftarrow$) separates the inferred Datalog
facts on the left-hand side from the existing Datalog fact(s) on the
right-hand side.
The first rule says the program-order relation implies the
must-happen-before relation.  The second rule says the
must-happen-before relation is transitive.

A Datalog solver, based on the above facts and rules, will
compute the maximal set of edges for the \textrel{HB} relation.
By sending a query to the Datalog solver, one may confirm that
$\textrel{HB}(s_1,s_5)$ indeed holds whereas $\textrel{HB}(s_2,s_3)$
does not hold.

\section{Constraint-based Synchronization Analysis}
\label{sec:constraint}

In this section, we present our method for computing abstract traces
of a single program.  In the next section, we leverage the abstract
traces of two programs to compute their differences.


First, we define the elementary relations that can be constructed
directly from the CFG of a program.

\vspace{1ex}
{
\begin{compactitem}
\item $\textrel{St}(s_1, th_1)$: Statement $s_1$ resides in Thread $th_1$
\item $\textrel{Po}(s_1, s_2)$: Statement $s_1$ is before $s_2$ in a thread
\item $\textrel{Dom}(s_1, s_2)$: Statement $s_1$ dominates $s_2$ in a thread 
\item $\textrel{PostDom}(s_1, s_2)$: $s_1$ post-dominates $s_2$ in a thread 
\item $\textrel{ThrdCreate}(th_1, s_1, th_2)$: Thread $th_1$ creates $th_2$ at $s_1$
\item $\textrel{ThrdJoin}(th_1, s_1, th_2)$: Thread $th_1$ joins back $th_2$ at $s_1$
\item $\textrel{CondWait}(s_1, v_1)$:  $s_1$  waits for condition variable $v_1$
\item $\textrel{CondSignal}(s_1, v_1)$: $s_1$ sends condition variable $v_1$
\item $\textrel{Load}(s_1, v_1)$: Statement $s_1$ reads from variable $v_1$
\item $\textrel{Store}(s_1, v_1)$: Statement $s_1$ writes to variable $v_1$
\item $\textrel{InCS}(s_1, l_1)$: $s_1$ resides in a critical section guarded by lock($l_1$)--unlock($l_1$) pair
\item $\textrel{SameCS}(s_1, s_2, l_1)$: $s_1$ and $s_2$ are in the same critical section guarded by $l_1$
\item $\textrel{DiffCS}(s_1, s_2, l_1)$: $s_1$ and $s_2$ are in different critical sections guarded by $l_1$
\end{compactitem}
}

While traversing the CFG to compute the \textrel{Po}, \textrel{Dom},
and \textrel{PostDom} relations, we take loops into consideration. For
example, two instructions involved with the same loop may not have
a \textrel{Dom} or \textrel{PostDom} relation, but an instruction
outside the loop can have a \textrel{Dom} or \textrel{PostDom}
relation with an instruction inside the loop.

Next, we define inference rules for computing new relations such
as \textrel{MayHb}, \textrel{MustHb}, and \textrel{MayRf}.

\subsection{Rules for Intra-thread Dependency}

To capture the execution order of instructions, we define the
following relations: $\textrel{MayHb}(s_1, s_2)$ means $s_1$ may happen
before $s_2$ in some execution, and $\textrel{MustHb}(s_1, s_2)$ means
$s_1$ happens before $s_2$ in all executions when both occur. 
Since the program order in each thread implies the execution order, we
have the following rule:
\begin{equation*}
{\footnotesize%
\begin{aligned}%
  \textrel{MustHb}(s_1, s_2) \leftarrow  \textrel{Po}(s_1, s_2)
\end{aligned}%
}%
\end{equation*}
In this work, we assume \emph{sequential consistency} but Datalog is capable of handling weaker memory models~\cite{KusanoW17} as well.

By definition \textrel{MustHb} implies \textrel{MayHb}, which means
\begin{equation*}
{\footnotesize%
\begin{aligned}%
  \textrel{MayHb}(s_1, s_2) \leftarrow  \textrel{MustHb}(s_1, s_2)
\end{aligned}%
}%
\end{equation*}

\subsection{Rules for Inter-thread Dependency}

When a parent thread $th_1$ creates a child thread $th_2$ at the statement $s_1$,
e.g., by invoking \texttt{pthread\_create}, any statement $s_2$ in
the child thread must occur after $s_1$.
\begin{equation*}
{\footnotesize%
\begin{aligned}%
\textrel{MustHb}(s_1, s_2) \leftarrow
    \textrel{ThrdCreate}(th_1, s_1, th_2) 
     \land \textrel{St}(s_2, th_2) 
\end{aligned}%
}%
\end{equation*}
Similarly, when a parent thread $th_1$ joins back a child thread $th_2$ at $s_1$, any
statement $s_2$ in $th_2$  must occur before $s_1$.
\begin{equation*}
{\footnotesize%
\begin{aligned}%
\textrel{MustHb}(s_2, s_1) \leftarrow 
    \textrel{ThrdJoin}(th_1, s_1, th_2) \land \textrel{St}(s_2, th_1) \\
\end{aligned}%
}%
\end{equation*}

\subsection{Rules for Signal-Wait Dependency}

When a condition variable $c$ is used, e.g., 
through \emph{signal(c)} and \emph{wait(c)}, it imposes an
execution order.
\begin{equation*}
{\footnotesize%
\begin{aligned}%
\textrel{MustHb}(s_1, s_2) \leftarrow 
    \textrel{CondSignal}(v_1, s_1) 
    \land \textrel{CondWait}(v_1, s_2) 
\end{aligned}%
}%
\end{equation*}

However, the rule needs to be used with caution.  In practice,
\emph{wait(c)} is often wrapped in an if-condition as
shown in Figure~\ref{fig:mot2-2}.  To be conservative, our method
analyzes the control flow of these threads and
applies the above rule only after detecting the usage pattern.
Since our method does not analyze the concrete values of any shared
variables, it does not check if the if-condition is valid.  Also,
developers may use condition variables in a different way.  Thus, in
our experiments (Section~\ref{sec:experiment}), we evaluated the
impact of this conservative approach---assuming the if-condition is
always valid---to confirm it does not lead to significant loss of
accuracy.

\subsection{Ad Hoc Synchronization}

\begin{figure}
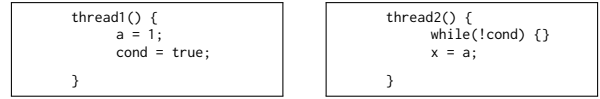

\vspace{1ex}
\centering
\begin{minipage}{.95\linewidth}
\centering
\framebox[.45\linewidth]{
\begin{minipage}{.45\linewidth}
{\scriptsize

\hspace{0.8cm}\texttt{thread1() \{}

\hspace{1.4cm}\texttt{a = 1;}

\hspace{1.4cm}\texttt{cond = true;}

\hspace{0.8cm}\texttt{\}}
}
\end{minipage}
}
\hspace{.05\linewidth}
\framebox[.45\linewidth]{
\begin{minipage}{.45\linewidth}
{\scriptsize

\hspace{0.8cm}\texttt{thread2() \{}

\hspace{1.4cm}\texttt{while(!cond) \{\}}

\hspace{1.4cm}\texttt{x = a;}

\hspace{0.8cm}\texttt{\}}
}
\end{minipage}
}
\vspace{1ex}
\end{minipage}

\vspace{-1ex}
\caption{\emph{Ad hoc} synchronization (\texttt{cond = false} initially).}
\label{fig:adhoc}
\vspace{-1ex}
\end{figure}

We handle \emph{ad hoc} synchronization similar to \emph{signal-wait}.
Fig.~\ref{fig:adhoc} shows an example where \texttt{cond} is a
user-added flag initialized to 0.  The busy-waiting 
in \texttt{thread2} ensures that \texttt{a=1} always occurs
before \texttt{x=a}.
By traversing the CFGs of these threads, we can identify the pattern;
this is practical since the number of usage patterns is
limited.
After that, we add a
\textrel{MustHb} edge from \texttt{cond=true}
to \texttt{while(!cond)}.
This is similar to adding \textrel{MustHb} edges for
\textrel{CondWait} and \textrel{CondSignal}.
As a result, we can decide the \emph{read-from} edge
between \texttt{x=a} and the initialization of \texttt{a} is
infeasible.

\subsection{Transitive Closure}

Since \textrel{MustHb} is transitive, we use the following rule to
compute the transitive closure:
\begin{equation*}
{\footnotesize%
\begin{aligned}%
  \textrel{MustHb}(s_1, s_3) \leftarrow  \textrel{MustHb}(s_1, s_2) \land 
\textrel{MustHb}(s_2, s_3) \\
\end{aligned}%
}%
\end{equation*}
When instructions in concurrent threads are not ordered
by \textrel{MustHb}, we assume they may occur in any order:
\begin{equation*}
{\footnotesize%
\begin{aligned}%
\textrel{MayHb}(s_1, s_2) \leftarrow 
    \textrel{St}(s_1, th_1) \land \textrel{St}(s_2, th_2)
    \land \lnot \textrel{MustHb}(s_2, s_1)\\
\end{aligned}%
}%
\end{equation*}
The \textrel{MayHb} relation is also transitive:
\begin{equation*}
{\footnotesize%
\begin{aligned}%
  \textrel{MayHb}(s_1, s_3) \leftarrow  \textrel{MayHb}(s_1, s_2) \land \textrel{MayHb}(s_2, s_3) \\
\end{aligned}%
}%
\end{equation*}

\subsection{Lock-enforced Critical Section}

For critical sections based on \emph{lock-unlock}, we introduce rules
based on access patterns.
First, we compute $\textrel{CoveredStore}(s_1, v_1, l_1)$,
meaning the store in $s_1$ is overwritten by a subsequent store in the
same critical section.  Consider
{\footnotesize \texttt{lk(a)} $\rightarrow$ \texttt{$W_1$(v)} $\rightarrow$ \texttt{$W_2$(v)} $\rightarrow$ \texttt{unlk(a)}}, 
where \texttt{$W_1$(v)} is a covered store and thus not visible to reads in other critical sections protected by the same lock.
\begin{equation*}
{\footnotesize%
\begin{aligned}%
  \textrel{CoveredStore}(s_1, v_1, l_1) 
   \leftarrow
     \textrel{Store}(s_1, v_1) 
     \land \textrel{Store}(s_2, v_1) \\
     \land \textrel{PostDom}(s_2, s_1)
     \land \textrel{SameCS}(s_1, s_2, l_1) \\ 
\end{aligned}%
}%
\end{equation*}
Similarly, \textrel{CoveredLoad}$(s_2,v_1,l_1)$ means the load
of $v_1$ in $s_2$ is covered and thus can only read from a preceding
store in the same critical section.
\begin{equation*}
{\footnotesize%
\begin{aligned}%
  \textrel{CoveredLoad}(s_2, v_1, l_1) 
   \leftarrow
     \textrel{Store}(s_1, v_1) 
     \land \textrel{Load}(s_2, v_1) \\
     \land \textrel{Dom}(s_1, s_2) 
     \land \textrel{SameCS}(s_1, s_2, l_1) \\ 
\end{aligned}%
}%
\end{equation*}
Consider {\footnotesize \texttt{lk(a)} $\rightarrow$ \texttt{W(v)}
$\rightarrow$ \texttt{R(v)} $\rightarrow$ \texttt{unlk(a)} } as an
example:  \texttt{R(v)} is covered by \texttt{W(v)} and thus
cannot read from stores in other critical sections protected by the same lock.

\subsection{Read-from Relation}

Finally, we compute \textrel{NoRf}$(s_1,s_2)$ which means the
read-from edge between $s_1$ and $s_2$ is infeasible.
\begin{equation*}
{\footnotesize%
\begin{aligned}%
  \textrel{NoRf}(s_1, s_2) 
   \leftarrow
     \textrel{Store}(s_1, v_1) 
     \land \textrel{Store}(s_3, v_1) 
     \land \textrel{Load}(s_2, v_1) \\
     \land \textrel{MustHb} (s_1, s_3) 
     \land \textrel{MustHb} (s_3, s_2) 
\end{aligned}%
}%
\end{equation*}
That is, in $\textrel{W}(x) \rightarrow \textrel{W}(x) \rightarrow \textrel{R}(x)$,
the first store  cannot be read by the load.
In addition to this generic rule, we have two more inference rules:
\begin{equation*}
{\footnotesize%
\begin{aligned}%
  \textrel{NoRf}(s_1, s_2) 
   \leftarrow
     \textrel{Store}(s_1, v_1) 
     \land \textrel{Load}(s_2, v_1) 
     \land \textrel{MayHb} (s_1, s_2) \\
     \land \textrel{CoveredLoad}(s_2, v_1, l_1) 
     \land \textrel{DiffCS}(s_1, s_2, l_1) \\ 
\end{aligned}%
}%
\end{equation*}
This rule means if one store may happen before one load, the load is
covered, and the store is in a different critical section,  the load
cannot read from the store.  This is because another store will
overwrite the value to be read.  
\begin{equation*}
{\footnotesize%
\begin{aligned}%
  \textrel{NoRf}(s_1, s_2) 
   \leftarrow
     \textrel{Store}(s_1, v_1) 
     \land \textrel{Load}(s_2, v_1) 
     \land \textrel{MayHb} (s_1, s_2) \\
     \land \textrel{CoveredStore}(s_1, v_1, l_1) 
     \land \textrel{DiffCS}(s_1, s_2, l_1) \\ 
\end{aligned}%
}%
\end{equation*}
This rule means if a store is covered, i.e., overwritten by a
subsequent store, the store cannot reach to any load in other critical sections protected by the
same lock.


We also compute \textrel{MayRf}$(s_1,s_2)$ which means the load in
$s_2$ may read from the store in $s_1$.
\begin{equation*}
{\footnotesize%
\begin{aligned}%
  \textrel{MayRf}(s_1, s_2) 
   \leftarrow
     \textrel{Store}(s_1, v_1) \land \textrel{Load}(s_2, v_1)  \land \textrel{MayHb}(s_1, s_2) \\
          \land \lnot \textrel{NoRf}(s_1, s_2)
\end{aligned}%
}%
\end{equation*}

\ignore{

And, \textrel{PostDom} and \textrel{Dom} are also transitive closure.

\begin{equation*}
{\footnotesize%
\begin{aligned}%
  \textrel{Dom}(s_1, s_3) \leftarrow  \textrel{Dom}(s_1, s_2) \land 
\textrel{Dom}(s_2, s_3) \\
  \textrel{PostDom}(s_1, s_3) \leftarrow  \textrel{PostDom}(s_1, s_2) \land 
\textrel{PostDom}(s_2, s_3) \\
\end{aligned}%
}%
\end{equation*}

}

\section{Computing the Differences}
\label{sec:optimization}

In this section, we show how to compare abstract traces of the two
programs to identify the differences.

\subsection{Symmetric Difference}
\label{sec:symmetric-diff}

\begin{figure}
\vspace{1ex}
\centering
\begin{minipage}{.90\linewidth}
\centering
\begin{minipage}{.45\linewidth}
\centering
\scalebox{.8}{
\begin{tikzpicture}
\coordinate (O) at (0,0);
\draw (O) node[fill=mypink,  circle, minimum size=1.5cm, draw, dashed] (A) {};
\draw (O) node[circle, minimum size=1cm, draw] (B) {$P_1$};
\draw (O) node[circle, minimum size=1.5cm, draw, color=blue, dashed] (C) {};

\begin{scope}[xshift=1.0cm]
\coordinate (O) at (0,0);
\draw (O) node[fill=white, circle, minimum size=1.5cm, draw, dashed] (D) {};
\draw (O) node[circle, minimum size=1cm, draw] (E) {$P_2$};
\draw (O) node[circle, minimum size=1.5cm, draw, color=blue, dashed] (F) {};
\end{scope}
\end{tikzpicture}
}
\end{minipage}
\begin{minipage}{.45\linewidth}
\centering
\scalebox{.8}{
\begin{tikzpicture}
\coordinate (O) at (1,0);
\draw (O) node[fill=mypink,  circle, minimum size=1.5cm, draw, dashed] (D) {};
\draw (O) node[circle, minimum size=1cm, draw] (E) {$P_2$};
\draw (O) node[circle, minimum size=1.5cm, draw, color=blue, dashed] (F) {};

\begin{scope}[xshift=0.0cm]
\coordinate (O) at (0,0);
\draw (O) node[fill=white, circle, minimum size=1.5cm, draw, dashed] (A) {};
\draw (O) node[circle, minimum size=1cm, draw] (B) {$P_1$};
\draw (O) node[circle, minimum size=1.5cm, draw, color=blue, dashed] (C) {};
\end{scope}
\end{tikzpicture}
}
\end{minipage}
\end{minipage}

\vspace{-1ex}
\caption{Differences of abstract traces:  $\Delta_{12}$ (left) and $\Delta{21}$ (right).}
\label{fig:diffDiagram}
\vspace{-1ex}
\end{figure}
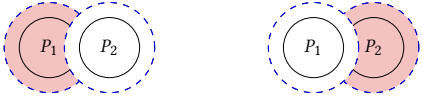

Fig.~\ref{fig:diffDiagram} shows the Venn diagram of our method for
computing the differences when given the abstract traces of two
programs.  The actual behaviors of programs $P_1$ and $P_2$ are
represented by the circles with solid lines.  The approximate
behaviors, in the form of abstract traces $\hat{T_1}$ and $\hat{T_2}$,
are represented by the circles with dashed lines.
Conceptually, the symmetric difference is computed based on
$\Delta_{12} = \hat{T_1}\setminus\hat{T_2}$ and $\Delta_{21}
=\hat{T_2}\setminus\hat{T_1}$, and for each is presented as pink-colored region in Fig.~\ref{fig:diffDiagram} (left and right).
The details of them are presented in the remainder of this section.

To compute the difference, we define two relations
\textrel{DiffP1} and \textrel{DiffP2} and rules for computing
them:
\begin{equation*}
{\footnotesize%
\begin{aligned}%
  \textrel{DiffP1}(s_1, s_2) \leftarrow \textrel{MayRf}(s_1, s_2, P_1) 
  \land \lnot \textrel{MayRf}(s_1, s_2, P_2) \\
  \textrel{DiffP2}(s_1, s_2) \leftarrow \textrel{MayRf}(s_1, s_2, P_2) 
  \land \lnot \textrel{MayRf}(s_1, s_2, P_1)
\end{aligned}%
}%
\end{equation*}
\textrel{DiffP1} represents edges that may happen 
in $P_1$ but not in $P_2$.  Similarly, \textrel{DiffP2}
represents edges that may happen in $P_2$ but not in  
$P_1$.  If \textrel{DiffP1} is not empty, there are more behaviors in
$P_1$; and if \textrel{DiffP2} is not empty, there are more behaviors
in $P_2$.

Since the Datalog solver may enumerate all
possible \textrel{MayHb} edges (used to compute \textrel{MayRf}), and the number of \textrel{MayHb}
edges increases rapidly as the program size increases, we need to
reduce the computational overhead.
Our insight is that, since we are only concerned with synchronization
differences in the end, as opposed to behaviors of the sequential computation, we
can restrict our analysis to instructions that
access global variables.  
Toward this end, we define a new relation named
$\textrel{Access}(v_1,s_1)$ which means $s_1$ accesses a global
variable $v_1$, and use it to guard the inference rules
for \textrel{MayHb} (and hence \textrel{MustHb}).  It forces the
Datalog solver to consider only global accesses, which reduces the
computational overhead without losing accuracy.  We demonstrate the
effectiveness of this optimization using experiments in
Section~\ref{sec:experiment}.

\subsection{Differences at Higher Ranks}

The rules so far use individual \emph{read-from} edges to characterize
the differences, which is equivalent to \emph{rank-1}
analysis~\cite{Bouajjani17}, but some programs may not have rank-1
difference but have differences of higher ranks.
To detect them, we need to compute \emph{ordered} sets of data-flow
edges allowed in one program but not in the other.

To be specific, for rank-2, we extend the \textrel{MayRf}
relation, which was defined over two instructions (an edge),
to \textrel{MayRfs} defined over four instructions, to represent an
ordered set of (two) \emph{read-from} edges.
Similarly, we extend the \textrel{NoRf} relation to \textrel{NoRfs},
which is also defined over four instructions to represent an ordered
set of (two) \emph{read-from} edges.

Previously, $\textrel{NoRf}(s_1,s_2)$ means there is no execution
trace where the store $s_1$ can be read by the load $s_2$, whereas
$\textrel{MayRf}(s_1,s_2)$ means there may exist some execution trace
that allows the \emph{read-from} edge $(s_1,s_2)$.
Similarly, $\textrel{NoRfs}((s_1,s_2), (s_3,s_4))$ means there is no
execution trace where the two \emph{read-from} edges $(s_1,s_2)$ and
$(s_3,s_4)$ occur together and in that order; and 
$\textrel{MayRfs}((s_1,s_2), (s_3,s_4))$ means there may exist some
execution trace that allows the two \emph{read-from} edges to occur
together and in that order.

\begin{figure}
\vspace{1ex}
\centering
\begin{minipage}{.95\linewidth}
\centering
\begin{minipage}{.4\linewidth}
\centering
\scalebox{.8}{
\begin{tikzpicture}[->, >=stealth',shorten >=1pt, auto, node distance=2.5 cm, 
                    semithick, scale=0.8, transform shape]

      \node[state]       (s1)  []   {$s_1$:W(x)};
      \node[state]       (s2)  [right=10mm of s1]   {$s_2$:W(x)};
      \node[state]       (s3)  [below=10mm of s1, xshift=13mm]   {$s_3$:R(x)};

      \path 
                (s1)  edge [left, xshift=-1mm] node       {\textbf{RF}}       (s3)
                (s2)  edge [right, xshift=1mm] node {\textbf{RF}}       (s3)

                ;

\end{tikzpicture}
}
\end{minipage}
\hspace{0.05\linewidth}
\begin{minipage}{.5\linewidth}
\centering
\scalebox{.8}{
\begin{tikzpicture}[->, >=stealth',shorten >=1pt, auto, node distance=2 cm, 
                    semithick, scale=0.8, transform shape]

      \node[state]       (s4)  [xshift=-10mm]   {$s_4$:R(x)};
      \node[state]       (s1)  [below=10mm of s4]   {$s_1$:W(x)};
      \node[state]       (s2)  [right=20mm of s4]   {$s_2$:R(x)};
      \node[state]       (s3)  [below=10mm of s2]   {$s_3$:W(x)};

      \path 
                (s1)  edge  node  [left,xshift=-3mm]     {\textbf{RF}}       (s2)
                (s3)  edge  node [right, xshift=3mm]   {\textbf{RF}}       (s4)
                (s2)  edge [right, dashed] node       {MustHB}       (s3)
                (s4)  edge [left, dashed]  node       {MustHB}       (s1)

                ;
\end{tikzpicture}
}
\end{minipage}
\end{minipage}

\vspace{-1ex}
  \caption{Illustrating the first two rank-2 inference rules.}
  \label{fig:rank2-rules-1}
\vspace{-1ex}
\end{figure}
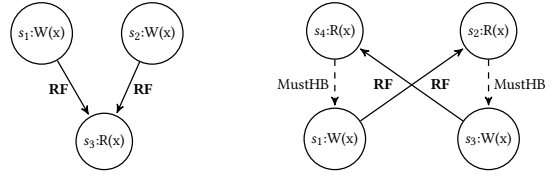

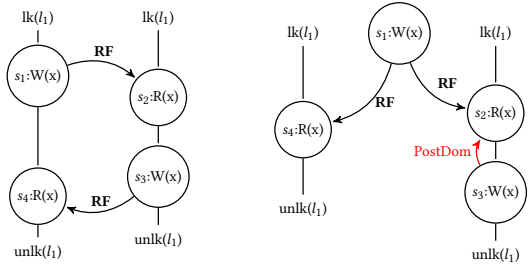
\begin{figure}
\vspace{1ex}
\centering
\begin{minipage}{.4\linewidth}
\centering
\scalebox{.8}{
\begin{tikzpicture}[->, >=stealth',shorten >=1pt, auto, node distance=2.5 cm, 
                    semithick, scale=0.8, transform shape]

      \node[draw=none]   (s10)                      {lk($l_1$)};
      \node[state]       (s1)  [below=3mm of s10]   {$s_1$:W(x)};
      \node[state]       (s4)  [below of=s1]        {$s_4$:R(x)};
      \node[draw=none]   (s40) [below=3mm of s4]    {unlk($l_1$)};

      \node[draw=none]   (s20) [right of=s10]       {lk($l_1$)};
      \node[state]       (s2)  [below=8mm of s20]   {$s_2$:R(x)};
      \node[state]       (s3)  [below=4mm of s2]    {$s_3$:W(x)};
      \node[draw=none]   (s30) [below=4mm of s3]    {unlk($l_1$)};

      \path 
                (s1)  edge [above,bend left] node       {\textbf{RF}}       (s2)
                (s3)  edge [left ,bend left] node[above]{\textbf{RF}}       (s4)

                (s10)  edge [-,below]            (s1)
                (s1)   edge [-,below]            (s4)
                (s4)   edge [-,below]            (s40)

                (s20)  edge [-,below]            (s2)
                (s2)   edge [-,below]            (s3)
                (s3)   edge [-,below]            (s30)

                ;

\end{tikzpicture}
}
\end{minipage}
\hspace{0.05\linewidth}
\begin{minipage}{.4\linewidth}
\centering
\scalebox{.8}{
\begin{tikzpicture}[->, >=stealth',shorten >=1pt, auto, node distance=2 cm, 
                    semithick, scale=0.8, transform shape]

      \node[draw=none]   (s04) []            {lk($l_1$)};
      \node[state]       (s4)  [below of=s04]   {$s_4$:R(x)};
      \node[draw=none]   (s40) [below=8mm of s4]    {unlk($l_1$)};

      \node[state]       (s1)  [right of=s04]       {$s_1$:W(x)};

      \node[draw=none]   (s20) [right of=s1]       {lk($l_1$)};
      \node[state]       (s2)  [below=8mm of s20]   {$s_2$:R(x)};
      \node[state]       (s3)  [below=4mm of s2]    {$s_3$:W(x)};
      \node[draw=none]   (s30) [below=4mm of s3]    {unlk($l_1$)};

      \path 
                (s1)  edge [bend right] node       {\textbf{RF}}       (s2)
                (s1)  edge [bend left]  node[right]{\textbf{RF}}       (s4)
                (s3)  edge [below,bend left,color=red]  node[left]{\textcolor{red}{PostDom}}       (s2)

                (s04)  edge [-,below]            (s4)
                (s4)   edge [-,below]            (s40)

                (s20)  edge [-,below]            (s2)
                (s2)   edge [-,below]            (s3)
                (s3)   edge [-,below]            (s30)

                ;
\end{tikzpicture}
}
\end{minipage}

\vspace{-1ex}
  \caption{Illustrating rank-2 rules related to lock-unlock.}
  \label{fig:rank2-rules-2}
\vspace{-1ex}
\end{figure}

First, we present our rules for computing $\textrel{NoRfs}$, which in
turn is used to compute $\textrel{MayRfs}$.  Since it is not possible
to enumerate all scenarios due to theoretical limitations, we resort
to the most common scenarios.  Nevertheless, we guarantee that
$\textrel{NoRfs}$ is an under-approximation, and the
corresponding $\textrel{MayRfs}$ is an over-approximation.
\begin{equation*}
{\footnotesize%
\begin{aligned}%
  \textrel{NoRfs}((s_1, s_3), (s_2, s_3))
   \leftarrow 
    \textrel{MayRf}(s_1, s_3) \land \textrel{MayRf}(s_2, s_3)
\end{aligned}%
}%
\end{equation*}
This rule is obvious because, as in Fig.~\ref{fig:rank2-rules-1}~(left), in the same execution trace a load
($s_3$) cannot read from two different stores ($s_1,s_2$).
\begin{equation*}
{\footnotesize%
\begin{aligned}%
\textrel{NoRfs}((s_1, s_2), (s_3, s_4)) \leftarrow  \textrel{MayRf}(s_1, s_2) \land \textrel{MayRf}(s_3, s_4)\\
     \land \textrel{MustHb}(s_2, s_3) \land \textrel{MustHb}(s_4, s_1) 
\end{aligned}%
}%
\end{equation*}
This rule is also obvious because, as shown in Fig.~\ref{fig:rank2-rules-1}~(right), if the two \emph{read-from} edges
form a cycle together with the must-happen-before edges, they lead to
a contradiction.

\begin{equation*}
{\footnotesize%
\begin{aligned}%
\textrel{NoRfs}((s_1, s_2), (s_3, s_4)) \leftarrow 
    \textrel{SameCS}(s_1, s_4, l_1)  \land \textrel{SameCS}(s_2, s_3, l_1) \\
    \land \textrel{DiffCS}(s_1, s_2, l_1)  
\end{aligned}%
}%
\end{equation*}
This rule is related to \emph{lock-unlock} pairs.  The rationale
behind it can be explained using the diagram in
Fig.~\ref{fig:rank2-rules-2}~(left).
Due to the \emph{lock-unlock} pairs, there are only two possible
interleavings: (1) if $s_1$ happens before $s_2$, $s_4$ must happen
before $s_3$ and $s_2$, which contradicts to the read-from edge
$(s_3,s_4)$; (2) if $s_3$ happens before $s_4$, $s_2$ must happen
before $s_1$, which contradicts to the read-from edge $(s_1,s_2)$.
Thus, the read-from edges cannot occur in the same execution trace.

Next, we define another rule related to \emph{lock-unlock} pairs. In
this rule, we use \textrel{PostDom}$(s_3,s_2)$ to mean, after $s_2$ is
executed, $s_3$ is guaranteed to be executed as well.
\begin{equation*}
{\footnotesize%
\begin{aligned}%
\textrel{NoRFs}((s_1, s_2), (s_1, s_4)) \leftarrow 
    \textrel{Store}(s_3, v_1) \land \textrel{PostDom}(s_3, s_2)  \\
    \land \textrel{DiffCS}(s_2, s_4, l_1) 
    \land \textrel{SameCS}(s_2, s_3, l_1)
\end{aligned}%
}%
\end{equation*}
The rationale behind this rule can be explained using the diagram in
Fig.~\ref{fig:rank2-rules-2}~(right).
Here, the loads and stores access the same variable.  If the read-from
edge $(s_1,s_2)$ is ahead of $(s_1,s_4)$ in the same execution trace, the
store in $s_3$ contradicts to the read-from edge $(s_1,s_4)$.


Finally, we compute \textrel{MayRFs} based on \textrel{NoRFs}:
\begin{equation*}
{\footnotesize%
\begin{aligned}%
\textrel{MayRFs}((s_1, s_2), (s_3, s_4)) \leftarrow  \lnot \textrel{NoRFs}( (s_1, s_2), (s_3, s_4) )
\end{aligned}%
}%
\end{equation*}
It means the read-from edges ($s_1,s_2$) and ($s_3,s_4$) may occur together and in
that order in some execution trace.
With \textrel{MayRFs}, we compute differences (\textrel{DiffP1}
and \textrel{DiffP2}) by replacing \textrel{MayRf}
with \textrel{MayRFs}.
Our method for computing differences of rank 3 or higher are similar,
and we omit the details for brevity.

\subsection{Example for Rank-2 Analysis}

Fig.~\ref{fig:mot3} shows an example that illustrates the rank-2
analysis.  Here, \texttt{thread1} sets \texttt{t} to 0 and \texttt{x}
to 1 before creating \texttt{thread2}.  Due to lock-unlock pairs,
the assertion cannot be violated in Fig.~\ref{fig:mot3-1}.
However, if the lock-unlock in \texttt{thread1} is removed as in
Fig.~\ref{fig:mot3-2}, the assertion may be violated because, in
between Lines~4 and 5, there may be a context switch which 
was not allowed previously.

However, the synchronization difference cannot be captured by
any individual \texttt{MayRF} edge.  In fact, the
table in Fig.~\ref{fig:motTable3} shows that the two programs have the
same set of \texttt{MayRF} edges.  In particular, since there are
two stores of \texttt{x}, the load at Line~2 may read from both Line~1
and Line~5.

To capture the difference, we need rank-2 analysis.  
\begin{itemize}
\item
Assume \texttt{RF(L1,L4)} occurs first,
meaning \texttt{thread2} acquires the lock and thus
prevents \texttt{thread1} from acquiring the same lock
until \texttt{thread2} exits the critical section.  It means the store
at Line~5 will set \texttt{x} to 2.  Therefore, the load of \texttt{x}
at Line~2 will have to read from Line~5, not from Line~1.  In other
words,
\texttt{RF(L1,L2)} cannot occur after \texttt{RF(L1,L4)} in the same execution.  
\item
Assume \texttt{RF(L1,L2)} occurs first and \texttt{thread2} will
not be executed until \texttt{thread1} finishes.  In this
case, \texttt{RF(L1,L4)} is allowed since no store of \texttt{x}
is in \texttt{thread1}.
\end{itemize}
As a result, the program in Fig.~\ref{fig:mot3-1} allows the ordered
set \{\texttt{RF(L1,L2)}, \texttt{RF(L1,L4)}\} but not the ordered
set \{\texttt{RF(L1,L4)}, \texttt{RF(L1,L2)}\}.

However, the program in Fig.~\ref{fig:mot3-2} allows the ordered
set \{\texttt{RF(L1,L4)}, \texttt{RF(L1,L2)}\} as well, due to the
removal of the lock-unlock pairs in \texttt{thread1}.
Specifically, when \texttt{RF(L1,L4)} occurs at the start of an execution, 
\texttt{thread1} may execute Line~2 before \texttt{thread2} execute  
Line~5, which allows Line~2 to read the value of \texttt{x} from
Line~1.

Our steps of conducting the rank-2 analysis, based on inference rules
presented so far, are shown in Fig.~\ref{fig:motTable3}.  There is no
difference in the \texttt{MayRF} sets; however, when comparing the
ordered set of \texttt{MayRF} edges, we can still see the difference. 
To support this analysis, we apply the aforementioned inference rules of
rank 2, which checks the existence of $(1,4) \rightarrow (1,2)$.

\begin{figure}
\vspace{1ex}
\centering

\subfigure[Before change]{
\centering
\begin{minipage}{.95\linewidth}
\centering
\framebox[.95\linewidth]{
\centering
\begin{minipage}{.95\linewidth}
\centering
\begin{tikzpicture}[node distance=4mm]
  \scriptsize

  \tikzstyle{node}=[minimum size=0pt]
  \tikzstyle{nnode}=[minimum size=0pt,inner sep=0pt]
  \tikzstyle{lnode}=[circle,draw,minimum size=4pt,inner sep=0pt,fill]

  \node[node] (x0) [] at (0, 0) {\texttt{thread1} \{};
  
  \node[node] (x1)  [below=0mm of x0, xshift=2mm] {\texttt{t = 0;}};
  
  \node[node] (x2)  [below=3mm of x0, xshift=0.3mm] {1:~~\texttt{x = 1;}};
     
  \node[node] (x3)  [below=6mm of x0, xshift=5.5mm] {\texttt{create(t2);}};
  
  \node[node] (x4)  [below=9mm of x0, xshift=3.5mm] {\texttt{lock(a);}};
  
  \node[node] (x5)  [below=13mm of x0, xshift=-1mm] {...};
    
  \node[node] (x6)  [below=15mm of x0, xshift=7mm] {2:~~\texttt{assert(x != t);}};
  
  \node[node] (x7)  [below=18mm of x0, xshift=5mm] {\texttt{unlock(a);}};
  
  \node[node] (x8)  [below=21mm of x0, xshift=-5.5mm] {\}};

  \node[node] (y0)  [right=25mm of x0] {\texttt{thread2} \{~~~~~~~};
  
  \node[node] (y1)  [below=1mm of y0, xshift=-3.7mm] {...};
  
  \node[node] (y2)  [below=3mm of y0, xshift=1mm] {\texttt{lock(a);}};
  
  \node[node] (y3)  [below=6mm of y0, xshift=-2.3mm] {4:~~\texttt{t = x;}};
  
  \node[node] (y4)  [below=10mm of y0, xshift=-3.7mm] {...};
  
  \node[node] (y5)  [below=12mm of y0, xshift=-2.3mm] {5:~~\texttt{x = 2;}};
  
  \node[node] (y6)  [below=15mm of y0, xshift=2.5mm] {\texttt{unlock(a);}};
  
  \node[node] (y7)  [below=19mm of y0, xshift=-3.7mm] {...};
  
  \node[node] (y8)  [below=21mm of y0, xshift=-8mm] {\}};

  \draw[->,>=stealth',thick,bend left] (x2.east) to node[draw=none,above, yshift=0.5mm]{\circled{1}:\textbf{RF}} (y3.west);
  
    \draw[->,>=stealth', dashed] (y3.west) to node[draw=none, yshift=-0.5mm]{} (x6.east); 

  \draw[->,>=stealth',thick] (y5.west) to node[draw=none,below, yshift=-0.5mm]{\circled{2}:\textbf{RF}} (x6.east);  
  
    \draw[->,>=stealth',thick,color=red, bend right, dashed] (x2.west) to node[draw=none,left, xshift=0mm]{\textcolor{red}{\textbf{RF}}} (x6.west);

\end{tikzpicture}

\end{minipage}

}
\vspace{1ex}
\end{minipage}

\label{fig:mot3-1}
}

\vspace{1ex}

\subfigure[After change]{
\centering
\begin{minipage}{.95\linewidth}
\centering
\framebox[.95\linewidth]{
\centering
\begin{minipage}{.95\linewidth}
\centering
\begin{tikzpicture}[node distance=4mm]
  \scriptsize

  \tikzstyle{node}=[minimum size=0pt]
  \tikzstyle{nnode}=[minimum size=0pt,inner sep=0pt]
  \tikzstyle{lnode}=[circle,draw,minimum size=4pt,inner sep=0pt,fill]

  \node[node] (x0) [] at (0, 0) {\texttt{thread1} \{};
  
  \node[node] (x1)  [below=0mm of x0, xshift=2mm] {\texttt{t = 0;}};
  
  \node[node] (x2)  [below=3mm of x0, xshift=0.3mm] {1:~~\texttt{x = 1;}};
     
  \node[node] (x3)  [below=6mm of x0, xshift=5.5mm] {\texttt{create(t2);}};
  
  \node[node] (x4)  [below=9mm of x0, xshift=3.5mm] {\st{\texttt{lock(a);}}};
  
  \node[node] (x5)  [below=13mm of x0, xshift=-1mm] {...};
  
  \node[node] (x8)  [below=15mm of x0, xshift=7mm] {2:~~\texttt{assert(x != t);}};  
  
  \node[node] (x7)  [below=18mm of x0, xshift=5mm] {\st{\texttt{unlock(a);}}};
  
  \node[node] (x9)  [below=21mm of x0, xshift=-5.5mm] {\}};

  \node[node] (y0)  [right=25mm of x0] {\texttt{thread2} \{~~~~~~~};
  
  \node[node] (y1)  [below=1mm of y0, xshift=-3.7mm] {...};
  
  \node[node] (y2)  [below=3mm of y0, xshift=1mm] {\texttt{lock(a);}};
  
  \node[node] (y3)  [below=6mm of y0, xshift=-2.3mm] {4:~~\texttt{t = x;}};
  
  \node[node] (y4)  [below=10mm of y0, xshift=-3.7mm] {...};
  
  \node[node] (y5)  [below=12mm of y0, xshift=-2.3mm] {5:~~\texttt{x = 2;}};
  
  \node[node] (y6)  [below=15mm of y0, xshift=2.5mm] {\texttt{unlock(a);}};
  
  \node[node] (y7)  [below=19mm of y0, xshift=-3.7mm] {...};
  
  \node[node] (y8)  [below=21mm of y0, xshift=-8mm] {\}};

  \draw[->,>=stealth',thick,bend left] (x2.east) to node[draw=none,above, yshift=0mm]{\circled{1}:\textbf{RF}} (y3.west);
  
    \draw[->,>=stealth', thick ] (y3.west) to node[draw=none,below, xshift=2mm]{\textbf{HB}} (x6.east);

    \draw[->,>=stealth',thick,color=red, bend right] (x2.west) to node[draw=none,left, xshift=0mm]{\textcolor{red}{\circled{2}:\textbf{RF}}} (x6.west);

\end{tikzpicture}

\end{minipage}

}
\vspace{1ex}
\end{minipage}
\label{fig:mot3-2}
}

\vspace{-1ex}
\caption{Example programs  with rank-2 differences.}
\label{fig:mot3}
\vspace{-1ex}
\end{figure}
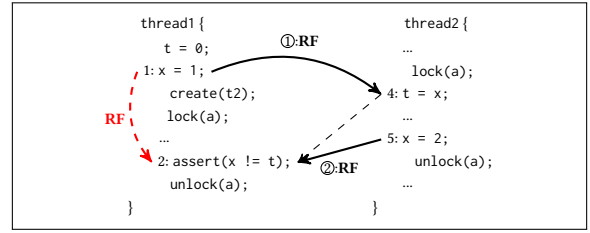
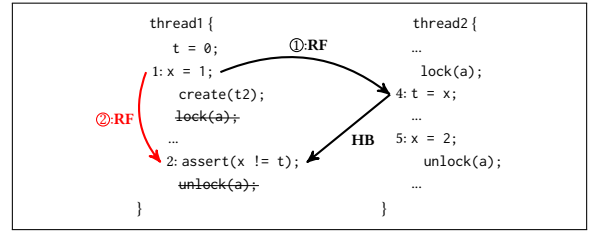

\begin{figure}
\centering

\subfigure{

\begin{minipage}{\linewidth}
\centering
\scalebox{0.8}{

 \begin{tabular}{ | c | c | c |}
 \hline
 
    & $mustHB$ & $ \{ (1, 2), (1,4), (1, 5), (4, 5) \}$ \\
    \cline{2-3}
    Fig~\ref{fig:mot3-1} & $mayHB$ & $mustHB$ $ \cup $ $ \{(2,4),(2,5),(4,2), (5,2) \} $ \\
    \cline{2-3}
     & MayRF & $ \{(1,2),(1,4), (5,2)\}$ \\
    \cline{2-3}
     & \textbf{Rank2} & $ \{[(1,2) \rightarrow (1,4)],[(1,4) \rightarrow (5,2)]\}$ \\
    \hline
    
    & $mustHB$ & $ \{ (1, 2), (1,4), (1, 5), (4, 5) \}$ \\
    \cline{2-3}
    Fig~\ref{fig:mot3-2} & $mayHB$ & $mustHB$ $ \cup $ $ \{(2,4),(2,5),(4,2), (5,2) \} $ \\
    \cline{2-3}
     & MayRF & $ \{(1,2),(1,4), (5,2)\}$ \\
    \cline{2-3}
     & \textbf{Rank2} & $ \{[(1,2) \rightarrow (1,4)],[(1,4) \rightarrow (5,2)],$ \\
     & & $ [\textbf{(1,4)} \rightarrow \textbf{(1,2)}] \} $  \\
    \hline

\end{tabular}
}

\vspace{1ex}
\end{minipage}
}

\vspace{-1ex}
\caption{Steps of our analysis for the programs in Fig.~\ref{fig:mot3}.}
\label{fig:motTable3}
\vspace{-2ex}
\end{figure}

\section{Experiments}
\label{sec:experiment}

We have implemented the method in a tool named \Name{}, which uses
LLVM~\cite{Adve03} as the frontend and $\mu Z$~\cite{Hoder11} in Z3 as
the Datalog solver at the backend.
Specifically, we use Clang/LLVM to parse the C/C++ code of
multithreaded programs and construct the LLVM intermediate
representation (IR).  Then, we traverse the LLVM IR to generate
program-specific Datalog facts.  These Datalog facts, when combined
with a set of program-independent inference rules, form the entire
Datalog program.  Finally, the $\mu Z$ Datalog solver is used to solve
the program, which repeatedly applies the rules to the fact until a
fixed point is reached.  By querying relations in the fixed
point, we can retrieve the analysis result.

\subsection{Experimental Setup}

We used two sets of benchmarks in our experiments.
The first set of benchmarks consists of 41 multithreaded programs,
which previously~\cite{Khoshnood15} have been used to illustrate
concurrency bug patterns found in real applications~\cite{Beyer15,
Bloem14, Lu08, Yu09, Yin11, llvm8441, gcc25530, gcc21334, gcc40518,
gcc3584, glib512624, jetty1187, Herlihy08}.  With these programs, our
goal is to evaluate how well the various types of concurrency bugs are
handled by our method, and how our results compare to that of the
prior technique based on model checking~\cite{Bouajjani17}.
For these benchmarks, the prior technique is not able to soundly instrument all applications.
Therefore, we manually insert assertions to be checked later by the CBMC
bounded model checker for detecting only one different edge.

The second set of benchmarks consists of 6 medium-sized applications
from open-source repositories; they have also been used
previously~\cite{Yang08, Yu09} to evaluate testing and automated
program repair tools.  Similarly, we are not able to
apply the prior technique~\cite{Bouajjani17} because it has limitations to instrument large size programs and it is impossible for us to manually insert assertions.
Nevertheless, we can evaluate how efficient our new method \Name{} is
on these real applications.
In total, our benchmarks has 13,500 lines of C code.

For each benchmark program, there are two versions, one of which is
the original program and the other is the changed program.  These
changed programs are patches collected from various sources: some are
from benchmarks used in prior research on 
testing~\cite{Yu09,Yang08} and repair~\cite{Khoshnood15}, whereas
others are from benchmarks used in differential analysis~\cite{Bouajjani17}.
We also created four programs, \emph{case1-4}, to illustrate
motivating examples used throughout this paper.
These benchmark programs, together with our experimental data, the
LLVM-based tool, as well as data obtained from applying the prior
technique~\cite{Bouajjani17}, have been made available online\footnote{https://github.com/ChunghaSung/EC-Diff}.

Our experiments were designed specifically to answer the following
research questions:
\begin{itemize}
\item 
Is our new method, based on a \emph{fast} and \emph{approximate}
static analysis as opposed to heavy-weight model checking techniques,
accurate enough for identifying the actual synchronization differences
in the benchmark programs?
\item 
Is our new method significantly more efficient, measured in terms of
the analysis time, than the prior technique based on model checking?
\end{itemize}
In all these experiments, we used a computer with an Intel Core
i5-4440 CPU @ 3.10 GHz x 4 CPUs with 12 GB of RAM, running the
Ubuntu-16.04 LTS operating system.




\subsection{Results on the First Set of Benchmarks}

Table~\ref{tbl:resultbox1} shows our results on the first set of
benchmarks, with 41 programs illustrating common bug patterns.
Columns~1 and 2 show the name and the number of lines of C code.
Column 3 shows the number of threads.  Column 4 shows the type of bug
illustrated by the program.  Specifically,
\texttt{Sync.} means the bug is due to misuse of locks, and thus to 
repair it, some \emph{lock-unlock} pairs have been added, removed or
modified;
\texttt{Cond.} means the bug is due to misuse of condition variables,
and thus to repair it, some \emph{signal-wait} pairs have been added,
removed or modified;
\texttt{Th.Order} means the bug is related to thread creation and join 
and thus involves \textrel{ThrdJoin} or \textrel{ThrdCreate}; and 
\texttt{Order.} means the bug is related to ordering of instructions 
imposed by ad-hoc synchronization.
Note that, in each of these benchmarks, there is some synchronization
difference.

The remaining columns show the statistics reported by \Name{} as well
as the prior technique~\cite{Bouajjani17}.
Specifically, Column 5 shows if \Name{} detected the synchronization
difference.  Column 6 shows at which rank our analysis is conducted
(Section~\ref{sec:optimization}): we iteratively increase the rank
starting from 1, until a synchronization difference is detected.  To
be efficient, we bound the rank to 3 during our evaluation.  
Columns 7 and 8 show the number of differences in $\Delta_{12}
= \hat{T_1} \setminus \hat{T_2}$ and $\Delta_{21}
= \hat{T_2} \setminus \hat{T_1}$.
For a rank-1 analysis, it is the number of read-from edges; for a
rank-2 or rank-3 analysis, it is the number of ordered sets of
read-from edges.
The next two columns show the total number
of \textrel{MayHb} edges (used to compute \textrel{MayRf}) in $P_1$ and $P_2$, respectively.

The last two columns compare the analysis time of our method
and the model checking time of the prior technique~\cite{Bouajjani17} to check one different edge.
%
%
%
For each benchmark, we limit the run time to one hour.

\begin{table*}
\vspace{1ex}
\caption{Experimental results on the first set of benchmark programs.}
\label{tbl:resultbox1}
\centering
\scalebox{0.8}{
\begin{tabular}{lccccccccccrr}
\toprule
 & & & &  \multicolumn{7}{c}{\Name{}} & ~~~~~Prior Technique~\cite{Bouajjani17} \\
\cmidrule(r){5-11}
\cmidrule(r){12-12}

Name & \ \ \ \ LoC\ \ \ \ & Threads & \ \ \ \ \ \ Type\ \ \ \ \ \  & Difference & Rank & $|\Delta_{12}|$ & $|\Delta_{21}|$ & \# of mayHB in $P_1$ & \# of mayHB in $P_2$ & \ \ Time (s)\ \  & Time (s)   \\
\midrule
case1                          &52  &3&Sync.   &yes&1&0&7   & 1,343& 1,343&0.26 & 11.53\\
case2                          &53  &3&Cond.   &yes&1&0&3   & 1,357& 1,474&0.26 &  4.80\\
case3                          &67  &3&Th.Order&yes&1&2&0   &   546&   482&0.19 & 46.64 \\
case4                          &94  &3&Sync.   &yes&2&0&1   &   421&   421&0.20 &  8.59 \\

\midrule
i2c-hid~\cite{Bouajjani17}     &76  &3&Sync.   &yes&1&1&0   & 2,570& 2,570&0.28 & 27.28 \\
i2c-hid-noa~\cite{Bouajjani17} &70  &3&Sync.   &yes&1&1&0   & 1,573& 1,573&0.26 &  7.48 \\
r8169-1~\cite{Bouajjani17}     &65  &3&Order   &yes&1&1&0   &   870&   852&0.25 &  3.38 \\
r8169-2~\cite{Bouajjani17}     &80  &3&Order   &yes&1&1&0   &   873&   839&0.25 &  2.17 \\
r8169-3~\cite{Bouajjani17}     &105 &4&Order   &yes&1&1&0   &   769&   769&0.25 &  8.37 \\
rtl8169-1~\cite{Bouajjani17}   &578 &8&Order   &yes&1&1&0   &60,741&60,691&0.89 & 1580.16 \\
rtl8169-2~\cite{Bouajjani17}   &578 &8&Order   &yes&1&1&0   &60,741&60,741&0.89 & 2384.14  \\
rtl8169-3~\cite{Bouajjani17}   &578 &8&Order   &no &3&0&0   &60,741&60,741&2.40 &  0.00 \\

\midrule
cherokee~\cite{Yu09}           &150 &3&Sync.   &yes&1&0&2   &1,148 & 1,148&0.31 & 7.59 \\
transmission~\cite{Yu09}       &91  &3&Cond.   &yes&1&1&0   &  690 &   613&0.29 & 6.89 \\
apache-21287~\cite{Yu09}       &74  &3&Sync.   &yes&1&2&0   &1,406 & 1,406&0.27 & 6.29\\
apache-25520~\cite{Yu09}       &181 &3&Sync.   &yes&2&8&0   &3,206 & 3,206&0.33 & 23.81\\

\midrule
account~\cite{Beyer15}         &82  &4&Cond.   &yes&1&0&2   &3,701 & 3,881&0.30 & 13.46 \\
barrier~\cite{Beyer15}         &138 &4&Cond.   &yes&1&3&0   &7,289 & 6,655&0.26 & 150.54 \\
boop~\cite{Beyer15}            &134 &3&Sync.   &yes&1&3&0   &2,625 & 2,625&0.25 & 8.90 \\
fibbench~\cite{Beyer15}        &63  &3&Cond.   &yes&1&0&71  &5,248 & 6,321&0.28 & 1483.33 \\
lazy~\cite{Beyer15}            &76  &4&Cond.   &yes&2&0&6   &3,409 & 3,549&0.24 & 32.16 \\
reorder~\cite{Beyer15}         &170 &5&Cond.   &yes&1&3&0   &9,493 & 8,737&0.40 & 12.79 \\
threadRW~\cite{Beyer15}        &147 &5&Cond.   &yes&1&2&0   &9,092 & 8,552&0.30 & 7.57 \\

\midrule
lineEq-2t~\cite{Bloem14}       &90  &3&Sync.   &yes&2&0&8   & 2,905& 2,905&0.30 & 23.34 \\
linux-iio~\cite{Bloem14}       &114 &3&Sync.   &yes&1&3&0   & 5,851& 5,851&0.31 & 24.13 \\
linux-tg3~\cite{Bloem14}       &130 &3&Cond.   &yes&1&2&0   &15,979&15,160&0.63 & 617.01 \\
vectPrime~\cite{Bloem14}       &127 &3&Sync.   &yes&2&2&0   &35,014&35,014&0.52 & 2.22\\

\midrule
mozilla-61369~\cite{Lu08}~~~~~~~ &84  &3&Cond.   &yes&1&0&1   &  473 &   565&0.25 & 3.57\\
mysql-3596~\cite{Lu08}         &92  &3&Cond.   &yes&1&1&0   &  773 &   733&0.25 & 3.82\\
mysql-644~\cite{Lu08}          &110 &3&Cond.   &yes&1&0&2   &1,343 & 1,434&0.33 & 5.40 \\

\midrule
counter-seq~\cite{Herlihy08}   &47  &3&Sync.   &yes&2&0&2   &1,135 & 1,135& 0.26  & 18.13 \\
ms-queue~\cite{Herlihy08}      &116 &3&Sync.   &yes&2&2&0   &5,754 & 5,754& 0.59  & 29.01 \\

\midrule
mysql5~\cite{Khoshnood15}      &59  &3&Sync.   &yes&2&0&4   &1,283 & 1,283& 0.20  & 22.92\\
freebsd-a~\cite{Yin11}         &176 &4&Cond.   &yes&1&0&22  &7,910 &10,109& 0.33 & 25.40 \\
llvm-8441~\cite{llvm8441}      &127 &3&Cond.   &yes&1&0&10  &3,042 & 3,118& 0.41 & 16.36\\
gcc-25530~\cite{gcc25530}      &87  &3&Sync.   &yes&2&2&0   &  806 &   806& 0.20  & 12.15 \\
gcc-3584~\cite{gcc3584}        &83  &3&Sync.   &yes&2&2&0   &1,843 & 1,843& 0.24  & 17.23 \\
gcc-21334~\cite{gcc21334}      &136 &3&Sync.   &yes&2&8&0   &5,290 & 5,290& 0.35  & 195.20 \\
gcc-40518~\cite{gcc40518}      &102 &3&Sync.   &yes&1&0&8   &3,027 & 3,027& 0.25 & 14.31 \\
glib-512624~\cite{glib512624}  &95  &3&Sync.   &yes&1&198&0 &5,748 & 5,748& 0.32 & $*>$3600.00\\
jetty-1187~\cite{jetty1187}    &69  &3&Sync.   &yes&2&0&2   &  885 &   885& 0.22  & 19.34 \\

\midrule
\textbf{Total}                 & & &        &   & &251&151 & 338,913 & 339,849 & 15.57 & $>$ 3h \\

\bottomrule

\end{tabular}
}

\begin{tablenotes}
      \item $*>3600.00$ means verification of the edge in $P_1$ succeeded, but verification of the edge in $P_2$ timed out after an hour. 
\end{tablenotes}

\end{table*}

Our results show \Name{} often finishes each benchmark in a second whereas the prior
technique can take up to 2,384 seconds (\emph{rtl8169-2}). In total, \Name{}
took less than 16 seconds whereas the prior technique took more than 3
hours.
In terms of accuracy, except for one program, \Name{} detected all
the synchronization differences.  This has been confirmed through
manual inspection where the reported differences are compared with
the ground truth.  Since we have randomly labeled the original and
changed programs as $P_1$ and $P_2$, some of the differences are in
$\Delta_{12}$ whereas the others are reported in $\Delta_{21}$.  In
total, \Name{} found 251 differences in $\Delta_{12}$ and 151
differences in $\Delta_{21}$.

The missed difference resides in \emph{rtl8169-3}: after running
the rank-3 analysis, our method still could not find it.  The reason
is because the differentiating behavior involves a deadlock and the
patch that removed it.  We explain why our method cannot detect it in
Section~\ref{sec:experiment-discussion}.

\subsection{Results on the Second Set of Benchmarks}

Table~\ref{tbl:resultbox2} shows our results on the second set of
benchmarks, consisting of six medium-sized programs. Note that these
programs are already out of the reach of the prior
technique~\cite{Bouajjani17} due to its requirement of manual code
instrumentation; therefore, we only report the statistics of
applying \Name{}.
Again, the original and modified programs are randomly labeled as
$P_1$ and $P_2$, respectively, to facilitate evaluation.

In total \Name{} found 30 differences in $\Delta_{12}$ and 42
differences in $\Delta_{21}$. Furthermore, all of them were found
during rank-1 analysis, and confirmed by manual inspection.
What is impressive is that these differences were identified
by sifting through a combined total of 24 million \textrel{MayHb}
edges, and yet, the analysis of all programs took only 140 seconds. 
The efficiency is, in large part, due to the restriction of our
analysis on instructions that access global variables as opposed to
all instructions in the program (refer to the last paragraph of
Section~\ref{sec:symmetric-diff}).
Otherwise, the number of \textrel{MayHb} edges would have been
orders-of-magnitude larger.


\begin{table*}
\vspace{1ex}
\caption{Experimental results on the second set of benchmark programs.}
\label{tbl:resultbox2}
\centering
\scalebox{0.8}{
\begin{tabular}{lccccccccccrrr}
\toprule
 & & & & \multicolumn{7}{c}{\Name{}}\\
\cmidrule(r){5-11}

Name\ \ \ \ \ \ \ \ \ \ \ \ \ \ \ \ \ \ \ \  & \ \ \ \ \ \ LoC\ \ \ \ \ \  & \ \ \ \ Threads\ \ \ \ & \ \ \ \ \ \ \ \ \ \ Type\ \ \ \ \ \ \ \ \ \ & Difference & \ \ Rank\ \ & \ \ \ \ $|\Delta_{12}|$\ \ \ \   & \ \ \ \ $|\Delta_{21}|$\ \ \ \   & \# of mayHB in $P_1$ & \# of mayHB in $P_2$ & \ \ \ \ Time (s)\ \ \ \ \\
\midrule

pbzip-1~\cite{Yu09, Yang08} & 1,143 &5 &Th.Order  &yes &1 &6   &0   &   782,846   &   773,934  & 14.98 \\
pbzip-2~\cite{Yu09, Yang08} & 1,143 &7 &Th.Order  &yes &1 &12  &0   & 1,150,404   & 1,135,428  & 30.61 \\
aget-1~\cite{Yu09, Yang08}  & 1,523 &4 &Cond.     &yes &1 &4   &0   & 1,099,047   & 1,078,695  &  9.41 \\
aget-2~\cite{Yu09, Yang08}  & 1,523 &6 &Cond.     &yes &1 &8   &0   & 3,218,034   & 3,162,684  & 28.60 \\
pfscan-1~\cite{Yang08}      & 1,327 &3 &Cond.     &yes &1 &0   &6   & 2,094,446   & 2,107,760  & 19.72 \\
pfscan-2~\cite{Yang08}      & 1,327 &5 &Cond.     &yes &1 &0   &36  & 4,138,361   & 4,164,989  & 39.96 \\
\midrule
\textbf{Total}              &        &  &          &    &  &30  &42  &12,483,138   & 1,242,3490 & 140.28 \\

\bottomrule

\end{tabular}
}

\end{table*}

\subsection{Discussion}
\label{sec:experiment-discussion}

Now, we answer the two research questions.  

\vspace{1ex}\noindent
\emph{Q1: Is \Name{} accurate enough for identifying 
synchronization differences?} 
The answer is yes.  As shown in our experimental results, \Name{}
produced a large number of differences, the majority of which are at
rank 1, which means they are individual \emph{read-from} edges allowed
in only one of the two programs, while the rest are at rank 2.
Although we do not guarantee that \Name{} finds all differentiating
behaviors, these detected ones have been confirmed by manual
inspection.

Given that these benchmarks contain  real concurrency bug patterns
reported and analyzed by many existing tools for testing and repair,
the result of \Name{} is sufficiently accurate.
The success in a large part is due to the nature of these programs,
where two versions behave almost same except for the thread
synchronization.  In such cases, our approximate analysis can come
really close to the ground truth.

\vspace{1ex}\noindent
\emph{Q2: Is \Name{} more efficient than the prior technique based on model checking?} 
The answer is yes.  As shown in our results, \Name{} was
10x to 1000x faster and, in total, completed differential analysis
of 13,500 lines of multithreaded C code in about 160 seconds.
In contrast, the prior technique took a longer time to analyze a
program.

Thus, we conclude that \Name{} is effective in identifying
synchronization differences in evolving programs. 
In practice, when developers update a program to fix concurrency bugs
or remove performance bugs (e.g., by
eliminating redundant locks), the differences in behavior are often
reflected in (sets of) data-flow edges being feasible in one version
but not in the other version.  Thus, computing these (sets of) data-flow edges
can be a fast way of checking if the changes introduce unexpected
behaviors.

\begin{figure}
\centering
\begin{minipage}{.95\linewidth}
\centering
\framebox[.4\linewidth]{
\begin{minipage}{.4\linewidth}
{\scriptsize

\hspace{0.8cm}\texttt{thread1() \{}

\hspace{1.4cm}\texttt{lock(a);}

\hspace{1.4cm}\texttt{lock(b);}

\hspace{1.4cm}\texttt{...}

\hspace{1.4cm}\texttt{unlock(b);}

\hspace{1.4cm}\texttt{unlock(a);}

\hspace{0.8cm}\texttt{\}}
}
\end{minipage}
}
\hspace{.17\linewidth}
\framebox[.4\linewidth]{
\begin{minipage}{.4\linewidth}
{\scriptsize

\hspace{0.8cm}\texttt{thread1() \{}

\hspace{1.4cm}\texttt{lock(b);}

\hspace{1.4cm}\texttt{lock(a);}

\hspace{1.4cm}\texttt{...}

\hspace{1.4cm}\texttt{unlock(a);}

\hspace{1.4cm}\texttt{unlock(b);}

\hspace{0.8cm}\texttt{\}}
}
\end{minipage}
}
\end{minipage}

\vspace{-1ex}
\caption{Code from \emph{rtl8169-3}: the original (left) and changed (right) versions.}
\label{fig:missing}
\vspace{-1ex}
\end{figure}

\emph{The Missing Case:}
Although \Name{} successfully detected most of the actual differences,
it missed the one in \emph{rtl8169-3}.  Fig.~\ref{fig:missing} shows
the code snippet of \texttt{thread1} from the original program ($P_1$)
on the left-hand side and the changed program
($P_2$) on the right-hand side.  The purpose of this patch is to
resolve a deadlock issue by changing the acquisition order of locks.
Since \Name{} focuses solely on data-flow edges, it is not able to
detect behavioral differences related to locking only.  In some
sense, this is a limitation shared by techniques relying on the
notion of abstract traces~\cite{ShashaS88,Bouajjani17}: the two
programs do not have data-related semantic difference other than the fact that
a deadlock exists in one program but does not exist in the other
program.

\section{Related Work}
\label{sec:relatedwork}

There has been prior work on statically computing the semantic
differences of sequential and concurrent programs.

For sequential programs, Jackson and Ladd~\cite{Jackson94} proposed a
method for computing the semantic differences by summarizing and
comparing the dependencies between input and output.
Godline and Strichman~\cite{Godlin10} proposed the use of inference
rules to prove the equivalence of two programs.  In the SymDiff
project, Lahiri et al.~\cite{Lahiri12,Lahiri13} developed a
language-agnostic assertion checking tool for computing the 
differences of imperative programs.
In the context of incremental symbolic execution~\cite{Person08},
various change-impact analysis techniques were used to identify
instructions that are affected by code modification and use the information to compute the
corresponding test inputs~\cite{Marinescu13}.
However, these methods are not directly applicable to
concurrent programs.


For concurrent programs, Joshi et al.~\cite{Joshi12} proposed the use
of failure frequencies of assertions to compare two programs, while
the general framework of refinement checking~\cite{Abadi91} could
also be applied to traces of two programs.  However, these techniques
are limited to individual executions.
Change-impact analysis~\cite{Lehnert11} were also applied to
concurrent programs, e.g., in regression testing~\cite{YuRothermel14},
prioritized scheduling~\cite{Jagannath11}, and incremental symbolic
execution~\cite{Guo16}.  However, these techniques focus on
reducing the cost of testing and analysis as opposed to identifying the
synchronization differences.

As we have mentioned earlier, the most closely related work is that of
Bouajjani et al.~\cite{Bouajjani17}, which computes the differences
between partial data-flow dependencies of two concurrent programs
using a bounded model checker.  However, the method is 
costly; furthermore, it requires code instrumentation to
insert assertions so they can be verified using a
model checker.  For example, it took about 30 minutes for a 
program (rtl8169) that can be analyzed by our method in less than a
second.  

Our method relies on the Datalog-based declarative program analysis
framework, which previously has been applied to both sequential and
concurrent programs as well as web applications~\cite{Heintze01,
Hajiyev06, Horwitz95, Bravenboer09, FarzanK12, Guo16, Guo15, Lam05,
Livshits05, Naik06, Whaley04, Sung16, ZhangMGNY14, AlbarghouthiKNS17}.
In the context of static analysis of concurrent programs, for example,
Kusano and Wang~\cite{KusanoW16, KusanoW17} used Datalog in a
thread-modular abstract interpretation to check the feasibility of
inter-thread data-flow edges on sequentially consistent and weaker
memory models.  Sung et al.~\cite{Sung17} used a similar technique for
modeling preemption scheduling of interrupts and thus improving the
accuracy of static analysis for interrupt-driven programs.  However,
none of these existing methods computes the synchronization
differences of evolving programs.

\section{Conclusions}
\label{sec:conclusion}

We have presented a \emph{fast} and \emph{approximate} static analysis
method for computing the synchronization differences of two concurrent
programs.  The method uses Datalog to capture structural information
of the programs, and uses a set of inference rules to codify the analysis
algorithm.  The analysis result, computed by an
off-the-shelf Datalog solver, consists of sets of data-flow edges
that are allowed by only one of the two programs.  We 
implemented the proposed method and evaluated it on a large number of
benchmark programs.  Our results show the method is
orders-of-magnitudes faster than the prior technique while being
sufficiently accurate in identifying the actual differences.  


\section*{Acknowledgments}
 
This work was supported in part by the U.S.\ National Science
Foundation (NSF) under grant CCF-1722710, the Office of Naval Research (ONR) under grant N00014-17-1-2896, the European Research Council (ERC) under the European Union's Horizon 2020 research and innovation program (grant agreement No 678177).
 
\clearpage\newpage
\bibliographystyle{plain} 
\bibliography{diffPrj}

\begin{thebibliography}{10}

\bibitem{gcc21334}
Gcc bug 21334.
\newblock \url {http://gcc.gnu.org/bugzilla/show\_bug.cgi?id=21334}.

\bibitem{gcc25530}
Gcc bug 24430.
\newblock \url {http://gcc.gnu.org/bugzilla/show\_bug.cgi?id=25330}.

\bibitem{gcc3584}
Gcc bug 3584.
\newblock \url {http://gcc.gnu.org/bugzilla/show\_bug.cgi?id=3584}.

\bibitem{gcc40518}
Gcc bug 40518.
\newblock \url {http://gcc.gnu.org/bugzilla/show\_bug.cgi?id=40518}.

\bibitem{glib512624}
Glib bug 51264.
\newblock \url {https://bugzilla.gnome.org/show\_bug.cgi?id=512624}.

\bibitem{jetty1187}
Jetty bug 1187.
\newblock \url {https://jira.codejaus.org/browse/JETTY-1187}.

\bibitem{llvm8441}
Llvm bug 8441.
\newblock \url {http://llvm.org/bugs/show\_bug.cgi?id=8441}.

\bibitem{Abadi91}
Mart\'{i}n Abadi and Leslie Lamport.
\newblock The existence of refinement mappings.
\newblock {\em Theoretical Computer Science}, 82(2):253--284, May 1991.

\bibitem{Adve03}
Vikram Adve, Chris Lattner, Michael Brukman, Anand Shukla, and Brian Gaeke.
\newblock {LLVA: A Low-level Virtual Instruction Set Architecture}.
\newblock In {\em {ACM/IEEE international symposium on Microarchitecture}}, Dec
  2003.

\bibitem{AlbarghouthiKNS17}
Aws Albarghouthi, Paraschos Koutris, Mayur Naik, and Calvin Smith.
\newblock Constraint-based synthesis of datalog programs.
\newblock In {\em International Conference on Principles and Practice of
  Constraint Programming}, pages 689--706, 2017.

\bibitem{Beyer15}
Dirk Beyer.
\newblock Software verification and verifiable witnesses.
\newblock In {\em International Conference on Tools and Algorithms for
  Construction and Analysis of Systems}, pages 401--416, 2015.

\bibitem{BindalBL13}
Sandeep Bindal, Sorav Bansal, and Akash Lal.
\newblock Variable and thread bounding for systematic testing of multithreaded
  programs.
\newblock In {\em International Symposium on Software Testing and Analysis},
  pages 145--155, 2013.

\bibitem{Bloem14}
Roderick Bloem, Georg Hofferek, Bettina K{\"o}nighofer, Robert K{\"o}nighofer,
  Simon Au{\ss}erlechner, and Raphael Sp{\"o}rk.
\newblock Synthesis of synchronization using uninterpreted functions.
\newblock In {\em International Conference on Formal Methods in Computer-Aided
  Design}, pages 11:35--11:42, 2014.

\bibitem{Bouajjani17}
Ahmed Bouajjani, Constantin Enea, and Shuvendu~K. Lahiri.
\newblock {\em Abstract Semantic Diffing of Evolving Concurrent Programs},
  pages 46--65.
\newblock Springer International Publishing, Cham, 2017.

\bibitem{Bravenboer09}
Martin Bravenboer and Yannis Smaragdakis.
\newblock Strictly declarative specification of sophisticated points-to
  analyses.
\newblock In {\em ACM SIGPLAN Conference on Object Oriented Programming,
  Systems, Languages, and Applications}, pages 243--262, 2009.

\bibitem{Dawson96}
Steven Dawson, C.~R. Ramakrishnan, and David~S. Warren.
\newblock Practical program analysis using general purpose logic programming
  systems\&mdash;a case study.
\newblock In {\em ACM SIGPLAN Conference on Programming Language Design and
  Implementation}, pages 117--126, 1996.

\bibitem{FarzanK12}
Azadeh Farzan and Zachary Kincaid.
\newblock Verification of parameterized concurrent programs by modular
  reasoning about data and control.
\newblock In {\em ACM SIGACT-SIGPLAN Symposium on Principles of Programming
  Languages}, pages 297--308, 2012.

\bibitem{Godlin10}
Benny Godlin and Ofer Strichman.
\newblock Time for verification.
\newblock chapter Inference Rules for Proving the Equivalence of Recursive
  Procedures, pages 167--184. Springer-Verlag, Berlin, Heidelberg, 2010.

\bibitem{Guo16}
Shengjian Guo, Markus Kusano, and Chao Wang.
\newblock {Conc-iSE}: Incremental symbolic execution of concurrent software.
\newblock In {\em IEEE/ACM International Conference On Automated Software
  Engineering}, pages 531--542, 2016.

\bibitem{Guo15}
Shengjian Guo, Markus Kusano, Chao Wang, Zijiang Yang, and Aarti Gupta.
\newblock Assertion guided symbolic execution of multithreaded programs.
\newblock In {\em ACM SIGSOFT Symposium on Foundations of Software
  Engineering}, pages 854--865, 2015.

\bibitem{Hajiyev06}
Elnar Hajiyev, Mathieu Verbaere, and Oege de~Moor.
\newblock {CodeQuest}: Scalable source code queries with datalog.
\newblock In {\em European Conference on Object-Oriented Programming}, pages
  2--27, 2006.

\bibitem{Heintze01}
Nevin Heintze and Olivier Tardieu.
\newblock Demand-driven pointer analysis.
\newblock In {\em ACM SIGPLAN Conference on Programming Language Design and
  Implementation}, pages 24--34, 2001.

\bibitem{Herlihy08}
Maurice Herlihy and Nir Shavit.
\newblock {\em The Art of Multiprocessor Programming}.
\newblock Morgan Kaufmann Publishers Inc., San Francisco, CA, USA, 2008.

\bibitem{Hoder11}
Krystof Hoder, Nikolaj Bj{\o}rner, and Leonardo de~Moura.
\newblock {muZ} - an efficient engine for fixed points with constraints.
\newblock In {\em International Conference on Computer Aided Verification},
  pages 457--462, 2011.

\bibitem{Horwitz95}
Susan Horwitz, Thomas Reps, and Mooly Sagiv.
\newblock Demand interprocedural dataflow analysis.
\newblock In {\em ACM SIGSOFT Symposium on Foundations of Software
  Engineering}, pages 104--115, 1995.

\bibitem{Jackson94}
Daniel Jackson and David~A. Ladd.
\newblock Semantic diff: A tool for summarizing the effects of modifications.
\newblock In {\em International Conference on Software Maintenance}, pages
  243--252, 1994.

\bibitem{Jagannath11}
Vilas Jagannath, Qingzhou Luo, and Darko Marinov.
\newblock Change-aware preemption prioritization.
\newblock In {\em International Symposium on Software Testing and Analysis},
  pages 133--143, 2011.

\bibitem{Joshi12}
Saurabh Joshi, Shuvendu~K. Lahiri, and Akash Lal.
\newblock Underspecified harnesses and interleaved bugs.
\newblock In {\em ACM SIGACT-SIGPLAN Symposium on Principles of Programming
  Languages}, pages 19--30, 2012.

\bibitem{Khoshnood15}
Sepideh Khoshnood, Markus Kusano, and Chao Wang.
\newblock Concbugassist: Constraint solving for diagnosis and repair of
  concurrency bugs.
\newblock In {\em International Symposium on Software Testing and Analysis},
  pages 165--176, 2015.

\bibitem{KusanoW16}
Markus Kusano and Chao Wang.
\newblock Flow-sensitive composition of thread-modular abstract interpretation.
\newblock In {\em ACM SIGSOFT Symposium on Foundations of Software
  Engineering}, pages 799--809, 2016.

\bibitem{KusanoW17}
Markus~J. Kusano and Chao Wang.
\newblock Thread-modular static analysis for relaxed memory models.
\newblock In {\em ACM SIGSOFT Symposium on Foundations of Software
  Engineering}, 2017.

\bibitem{Lahiri12}
Shuvendu~K. Lahiri, Chris Hawblitzel, Ming Kawaguchi, and Henrique Reb\^{e}lo.
\newblock {SYMDIFF}: A language-agnostic semantic diff tool for imperative
  programs.
\newblock In {\em International Conference on Computer Aided Verification},
  pages 712--717, 2012.

\bibitem{Lahiri13}
Shuvendu~K. Lahiri, Kenneth~L. McMillan, Rahul Sharma, and Chris Hawblitzel.
\newblock Differential assertion checking.
\newblock In {\em ACM SIGSOFT Symposium on Foundations of Software
  Engineering}, pages 345--355, 2013.

\bibitem{Lam05}
Monica~S. Lam, John Whaley, V.~Benjamin Livshits, Michael~C. Martin, Dzintars
  Avots, Michael Carbin, and Christopher Unkel.
\newblock Context-sensitive program analysis as database queries.
\newblock In {\em ACM SIGMOD-SIGACT-SIGART Symposium on Principles of Database
  Systems}, pages 1--12, 2005.

\bibitem{Lehnert11}
Steffen Lehnert.
\newblock A taxonomy for software change impact analysis.
\newblock In {\em International Workshop on Principles of Software Evolution
  and Annual ERCIM Workshop on Software Evolution}, pages 41--50, 2011.

\bibitem{Livshits05}
V.~Benjamin Livshits and Monica~S. Lam.
\newblock Finding security vulnerabilities in java applications with static
  analysis.
\newblock In {\em USENIX Security Symposium}, pages 18--18, 2005.

\bibitem{Lu08}
Shan Lu, Soyeon Park, Eunsoo Seo, and Yuanyuan Zhou.
\newblock Learning from mistakes: A comprehensive study on real world
  concurrency bug characteristics.
\newblock In {\em International Conference on Architectural Support for
  Programming Languages and Operating Systems}, pages 329--339, 2008.

\bibitem{Marinescu13}
Paul~Dan Marinescu and Cristian Cadar.
\newblock {KATCH}: High-coverage testing of software patches.
\newblock In {\em ACM SIGSOFT Symposium on Foundations of Software
  Engineering}, pages 235--245, 2013.

\bibitem{MusuvathiQBBNN08}
Madanlal Musuvathi, Shaz Qadeer, Thomas Ball, G{\'e}rard Basler,
  Piramanayagam~Arumuga Nainar, and Iulian Neamtiu.
\newblock Finding and reproducing heisenbugs in concurrent programs.
\newblock In {\em USENIX Symposium on Operating Systems Design and
  Implementation}, pages 267--280, 2008.

\bibitem{Naik06}
Mayur Naik, Alex Aiken, and John Whaley.
\newblock Effective static race detection for java.
\newblock In {\em ACM SIGPLAN Conference on Programming Language Design and
  Implementation}, pages 308--319, 2006.

\bibitem{Person08}
Suzette Person, Matthew~B. Dwyer, Sebastian Elbaum, and Corina~S.
  P\v{a}s\v{a}reanu.
\newblock Differential symbolic execution.
\newblock In {\em ACM SIGSOFT Symposium on Foundations of Software
  Engineering}, pages 226--237, 2008.

\bibitem{Ramalingam00}
G.~Ramalingam.
\newblock Context-sensitive synchronization-sensitive analysis is undecidable.
\newblock {\em ACM Trans. Program. Lang. Syst.}, 22(2):416--430, 2000.

\bibitem{ShashaS88}
Dennis Shasha and Marc Snir.
\newblock Efficient and correct execution of parallel programs that share
  memory.
\newblock {\em {ACM} Trans. Program. Lang. Syst.}, 10(2):282--312, 1988.

\bibitem{Sung16}
Chungha Sung, Markus Kusano, Nishant Sinha, and Chao Wang.
\newblock Static {DOM} event dependency analysis for testing web applications.
\newblock In {\em ACM SIGSOFT Symposium on Foundations of Software
  Engineering}, pages 447--459, 2016.

\bibitem{Sung17}
Chungha Sung, Markus Kusano, and Chao Wang.
\newblock Modular verification of interrupt-driven software.
\newblock In {\em IEEE/ACM International Conference On Automated Software
  Engineering}, pages 206--216, 2017.

\bibitem{WangYGG08}
Chao Wang, Yu~Yang, Aarti Gupta, and Ganesh Gopalakrishnan.
\newblock Dynamic model checking with property driven pruning to detect race
  conditions.
\newblock In {\em International Symposium on Automated Technology for
  Verification and Analysis}, pages 126--140, 2008.

\bibitem{Whaley04}
John Whaley and Monica~S. Lam.
\newblock Cloning-based context-sensitive pointer alias analysis using binary
  decision diagrams.
\newblock In {\em ACM SIGPLAN Conference on Programming Language Design and
  Implementation}, pages 131--144, 2004.

\bibitem{Yang08}
Yu~Yang, Xiaofang Chen, and Ganesh Gopalakrishnan.
\newblock Inspect: A runtime model checker for multithreaded {C} programs.
\newblock Technical report, University of Utah, 2008.

\bibitem{YangCGW09}
Yu~Yang, Xiaofang Chen, Ganesh Gopalakrishnan, and Chao Wang.
\newblock Automatic discovery of transition symmetry in multithreaded programs
  using dynamic analysis.
\newblock In {\em International SPIN Workshop on Model Checking Software},
  pages 279--295, 2009.

\bibitem{Yin11}
Zuoning Yin, Ding Yuan, Yuanyuan Zhou, Shankar Pasupathy, and Lakshmi
  Bairavasundaram.
\newblock How do fixes become bugs?
\newblock In {\em ACM SIGSOFT Symposium and the 13th European Conference on
  Foundations of Software Engineering}, pages 26--36, 2011.

\bibitem{Yu09}
Jie Yu and Satish Narayanasamy.
\newblock A case for an interleaving constrained shared-memory multi-processor.
\newblock In {\em International Symposium on Computer Architecture}, pages
  325--336, 2009.

\bibitem{YuRothermel14}
Tingting Yu, Witawas Srisa-an, and Gregg Rothermel.
\newblock {SimRT}: An automated framework to support regression testing for
  data races.
\newblock In {\em International Conference on Software Engineering}, pages
  48--59, 2014.

\bibitem{ZhangMGNY14}
Xin Zhang, Ravi Mangal, Radu Grigore, Mayur Naik, and Hongseok Yang.
\newblock On abstraction refinement for program analyses in datalog.
\newblock In {\em ACM SIGPLAN Conference on Programming Language Design and
  Implementation}, pages 239--248, 2014.

\end{thebibliography}

\end{document}